\documentclass[a4paper,fleqn,usenatbib]{mnras}

\usepackage{amsmath}	
\usepackage{amssymb}	
\usepackage{mathptmx}
\usepackage{txfonts}

\usepackage[T1]{fontenc}
\usepackage{ae,aecompl}

\usepackage{graphicx}	
\usepackage{pdflscape} 
\usepackage{longtable} 
\usepackage{color}

\newcommand{\hb}{H$\beta$}

\newcommand{\fci}{[C\thinspace\textsc{i}]}
\newcommand{\foiii}{[O\thinspace\textsc{iii}]}

\newcommand{\foi}{[O\thinspace\textsc{i}]}
\newcommand{\foii}{[O\thinspace\textsc{ii}]}
\newcommand{\fsii}{[S\thinspace\textsc{ii}]}
\newcommand{\fsiii}{[S\thinspace\textsc{iii}]}
\newcommand{\fni}{[N\thinspace\textsc{i}]}
\newcommand{\fnii}{[N\thinspace\textsc{ii}]}
\newcommand{\fariv}{[Ar\thinspace\textsc{iv}]}
\newcommand{\fcliii}{[Cl\thinspace\textsc{iii}]}

\newcommand{\fneiii}{[Ne\thinspace\textsc{iii}]}
\newcommand{\ffeii}{[Fe\thinspace\textsc{ii}]}
\newcommand{\ffeiii}{[Fe\thinspace\textsc{iii}]}

\newcommand{\fniqii}{[Ni\thinspace\textsc{ii}]}

\newcommand{\oiii}{O\thinspace\textsc{iii}}

\newcommand{\nii}{N\thinspace\textsc{ii}}

\newcommand{\oi}{O\thinspace\textsc{i}}
\newcommand{\oii}{O\thinspace\textsc{ii}}

\newcommand{\cii}{C\thinspace\textsc{ii}}
\newcommand{\ciii}{C\thinspace\textsc{iii}}

\newcommand{\neiii}{Ne\thinspace\textsc{iii}}
\newcommand{\sii}{S\thinspace\textsc{ii}}
\newcommand{\siii}{S\thinspace\textsc{iii}}

\newcommand{\cliii}{Cl\thinspace\textsc{iii}}
\newcommand{\fariii}{[Ar\thinspace\textsc{iii}]}

\newcommand{\feii}{Fe\thinspace\textsc{ii}}
\newcommand{\feiii}{Fe\thinspace\textsc{iii}}

\newcommand{\ariii}{Ar\thinspace\textsc{iii}}

\newcommand{\hi}{H\,\textsc{i}}
\newcommand{\hii}{H\thinspace\textsc{ii}}

\newcommand{\hei}{He\thinspace\textsc{i}}
\newcommand{\heii}{He\thinspace\textsc{ii}}
\newcommand{\mgi}{Mg\thinspace\textsc{i}}
\newcommand{\SIii}{Si\thinspace\textsc{ii}}

\title[C and O abundances in NGC\,300 and M\,33]{Carbon and oxygen abundance gradients in NGC\,300 and M\,33 from optical recombination lines\thanks{Based on observations collected at the European Southern Observatory, Chile, proposal number ESO 085.B-0072(A) and the Gran Telescopio Canarias (GTC), instaled in the Spanish Observatorio del Roque de los Muchachos of the Instituto de Astrof\'isica de Canarias, in the island of La Palma, programme 6-GTC1/13B.}}
\author[L.~Toribio San Cipriano et al.]
	{L.~Toribio San Cipriano,$^{1,2}$\thanks{E-mail:ltoribio@iac.} 
	J.~Garc\'ia-Rojas,$^{1,2}$ C.~Esteban,$^{1,2}$
	\newauthor 
	F.~Bresolin,$^{3}$ M.~Peimbert$^{4}$\\
	$^{1}$Instituto de Astrof\'isica de Canarias, E-38200, La Laguna, Tenerife, Spain\\
	$^{2}$Departamento de Astrof\'isica, Universidad de La Laguna, E-38206, La Laguna, Tenerife, Spain\\
	$^{3}$Institute for Astronomy, 2680 Woodlawn Drive, Honolulu, HI 96822, USA\\
	$^{4}$Instituto de Astronom\'\i a, UNAM, Apdo. Postal 70-264, 04510 M\'exico D.F., Mexico\\
}

\date{Accepted 2016 February 17. Received 2016 February 17; in original form 2016 January 19}

\pubyear{2015}

\begin{document}

\label{firstpage}
\pagerange{\pageref{firstpage}--\pageref{lastpage}} 
\maketitle

\begin{abstract}
We present deep spectrophotometry of several {\hii} regions in the nearby low-mass spiral galaxies NGC~300 and M\,33. The data have been taken with UVES and OSIRIS spectrographs attached to the 8\,m VLT and 10.4\,m GTC telescopes, respectively. We have derived precise values of the physical conditions for each object making use of several emission line-intensity ratios. In particular, we have obtained direct determinations of the electron temperature in all the observed objects. We detect pure recombination lines (RLs) of {\cii} and {\oii} in several of the {\hii} regions, permitting to derive their C/H and C/O ratios. We have derived the radial abundance gradient of O for each galaxy making use of collisionally excited lines (CELs) and RLs, as well as the C and N gradients using RLs and CELs, respectively. We obtain the first determination of the C/H gradient of NGC~300 and improve its determination in the case of M\,33. In both galaxies, the C/H gradients are steeper that those of O/H, leading to negative C/O gradients. Comparing with similar results for other spiral galaxies, we find a strong correlation between the slope of the C/H gradient and $M_\mathrm{V}$. We find that some {\hii} regions located close to the isophotal radius ($R_{25}$) of NGC~300 and M\,33 show C/O ratios more similar to those typical of dwarf galaxies than those of  {\hii} regions in the discs of more massive spirals. This may be related to the absence of flattening of the gradients in the external parts of NGC~300 and M\,33. Finally, we find very similar N/H gradients in both galaxies and a fair correlation between the slope of the N/H gradient and $M_\mathrm{V}$ comparing with similar data for a sample of spiral galaxies. 

\end{abstract}


\section{Introduction}

The analysis of emission-line spectra in \hii\ regions has been used for decades to trace the chemical composition of the ionized gas phase of the interstellar medium (ISM). The knowledge of the chemical composition across galactic discs is crucial to understand nuclear processes in stellar interiors and the formation and chemical evolution of galaxies.

The analysis of collisionally excited lines (hereafter CELs) in the spectra of \hii\ regions provide chemical abundances of elements such as N, O, Ne, S, Cl, Ar and Fe. Carbon (C) is the second most abundant heavy element in the Universe and it is of paramount biogenic importance, in addition it is an important source of opacity and energy production in stars as well as a major constituent of interstellar dust and organic molecules. Notwithstanding, C abundance determinations in extragalactic \hii\ regions are scarce in the literature. Almost all these C abundances are derived from \ciii] 1909 \AA\ and \cii] 2326 \AA\  CELs in the UV, which can only be observed from space and are strongly affected by the uncertainty in the choice of UV reddening function \citep{1995ApJ...443...64G, 1999ApJ...513..168G}. On the other hand, the emission of the far-IR [\cii] 158 $\mu$m fine-structure line appears predominantly in photodissociation regions, not in the ionized gas-phase. However, thanks to the new large aperture telescopes is possible to determine C abundances using the faint optical recombination lines (hereafter RLs) in bright nebulae. Some authors \citep[e.g.,][]{1992RMxAA..24..155P, 2002ApJ...581..241E, 2009ApJ...700..654E, 2014MNRAS.443..624E, 2007ApJ...670..457G, 2007ApJ...656..168L} measured the \cii\ 4267\AA\ RL in Galactic and extragalactic \hii\ regions. Besides they could also measure the faint RLs of the multiplet 1 of \oii\ at about 4650 \AA. The calculation of chemical abundances through RLs allows the study of the behaviour of C/H and C/O across the disc of different galaxies as well as to better understand the nucleosynthetic origin of the carbon and its enrichment timescale \citep{2005ApJ...623..213C}. In this paper we aim to study the behaviour of the C/O ratio in two nearby spiral galaxies: NGC\,300 and M\,33.

The proximity of the spiral galaxy NGC\,300 (1.88 Mpc) and its relative low inclination ($\mathrm{i} = 39^{\mathrm{o}}.8$) makes it a perfect candidate for numerous investigations. \citet{1979MNRAS.189...95P} made the first chemical abundances analysis in NGC\,300 using \hii\ regions. Since then many other authors increased the number of such studies \citep{1983MNRAS.204..743W, 1984MNRAS.211..507E, 1988A&AS...73..407D, 1997A&A...322...41C}. However, despite of all these works, only two \hii\ regions of NGC\,300 had chemical abundances derived via the ``direct" method, i. e. using the \foiii\ $\lambda$4363 line to determine the electron temperature. The work of \citet{2009ApJ...700..309B} about \hii\ regions in NGC\,300 yielded a significant improvement in the study of gas-phase metallicities in this galaxy because they measured the \foiii\ $\lambda$4363 line in 28 \hii\ regions.

M\,33 (NGC\,598) is the third spiral galaxy in size within the Local Group, located at 840 kpc \citep{1991ApJ...372..455F} with an intermediate inclination $\mathrm{i} = 53^{\mathrm{o}}$ \citep{1989AJ.....97...97Z}. \citet{1971ApJ...168..327S} was the first to determine radial abundance gradients in this galaxy. Today, we can find many studies about radial abundance gradients in M\,33 derived from \hii\ regions in the literature \citep[e.g.][]{1988MNRAS.235..633V, 2006ApJ...637..741C, 2007AA...470..865M, 2008ApJ...675.1213R, 2011ApJ...730..129B}.
\citet{2008ApJ...675.1213R} determined a value for the O gradient of $8.35 (\pm 0.04) -0.18 (\pm 0.08) R/R_{25}$ based on detections of the \foiii 4363 \AA\ line. In addition, they found an intrinsic scatter of 0.11 dex around the gradient which is higher than the measured uncertainties. With the goal of verifying this local dispersion in the oxygen abundances, \citet{2011ApJ...730..129B} studied 25 \hii\ regions in the inner part of M\,33. They found a much lower scatter (around 0.06 dex) than that observed by \citet{2008ApJ...675.1213R} which can be explained by the different measuring uncertainties. The radial oxygen abundance gradient measured by \citet{2011ApJ...730..129B} was $8.48 (\pm 0.04) -0.29 (\pm 0.07) R/R_{25}$. The slope of this gradient is larger than that determined by \citet{2008ApJ...675.1213R} but is in good agreement with the gradient obtained by \citet{2007AA...470..865M}, which is $8.53 (\pm 0.05) -0.36 (\pm 0.07) R/R_{25}$.

Traditionaly, the standard method for deriving ionic abundances in \hii\ regions is based in the intensity of bright CELs because they are much easier to detect than RLs. However, the intensity of RLs is less dependent on the precision of the electron temperature determination, providing -- in principle -- more stable abundance values. Unfortunately, abundances based on RLs are sistematically higher than those derived from CELs. This is called the ``abundances discrepancy" problem which has been controversial for more than 40 years \citep{1969BOTT....5....3P, 2005MNRAS.364..687T, 2012ApJ...752..148N}. The measurement of \oii\ RLs allows us to compare  abundance values of $\mathrm{O}^{++}$ obtained with both kinds of lines (CELs and RLs). This is the first work where the $\mathrm{C}^{++}$ abundances are determined from \cii\ RLs in NGC\,300 and, in addition, we double the number of \hii\ regions with C abundances derived from RLs in M\,33 --\citet{2002ApJ...581..241E, 2009ApJ...700..654E} determined previously the C/H ratio in NGC\,595 and NGC\,604.

This paper is organized as follows: in Sections \ref{sec:obs} and \ref{sec:iden} we describe the observations, the data reduction procedure, the measurement of the emission lines and the determination of reddening coefficients. In Section \ref{sec:physicalcondition} we determine the electron temperature and density of the ionized gas. In Section \ref{sec:abundances} we present the ionic abundances from CELs and RLs and the total abundances. In Section \ref{sec:discusion} we discuss the radial C, N and O abundance gradients and compare the nebular O abundances with stellar determinations. We summarize our main results in Section \ref{sec:conclusions}.

\section{Observations and data reduction}\label{sec:obs}

The observational data acquisition and the reduction process for each galaxy is described below.

\begin{figure*}
  \centering
   \includegraphics[scale=0.39]{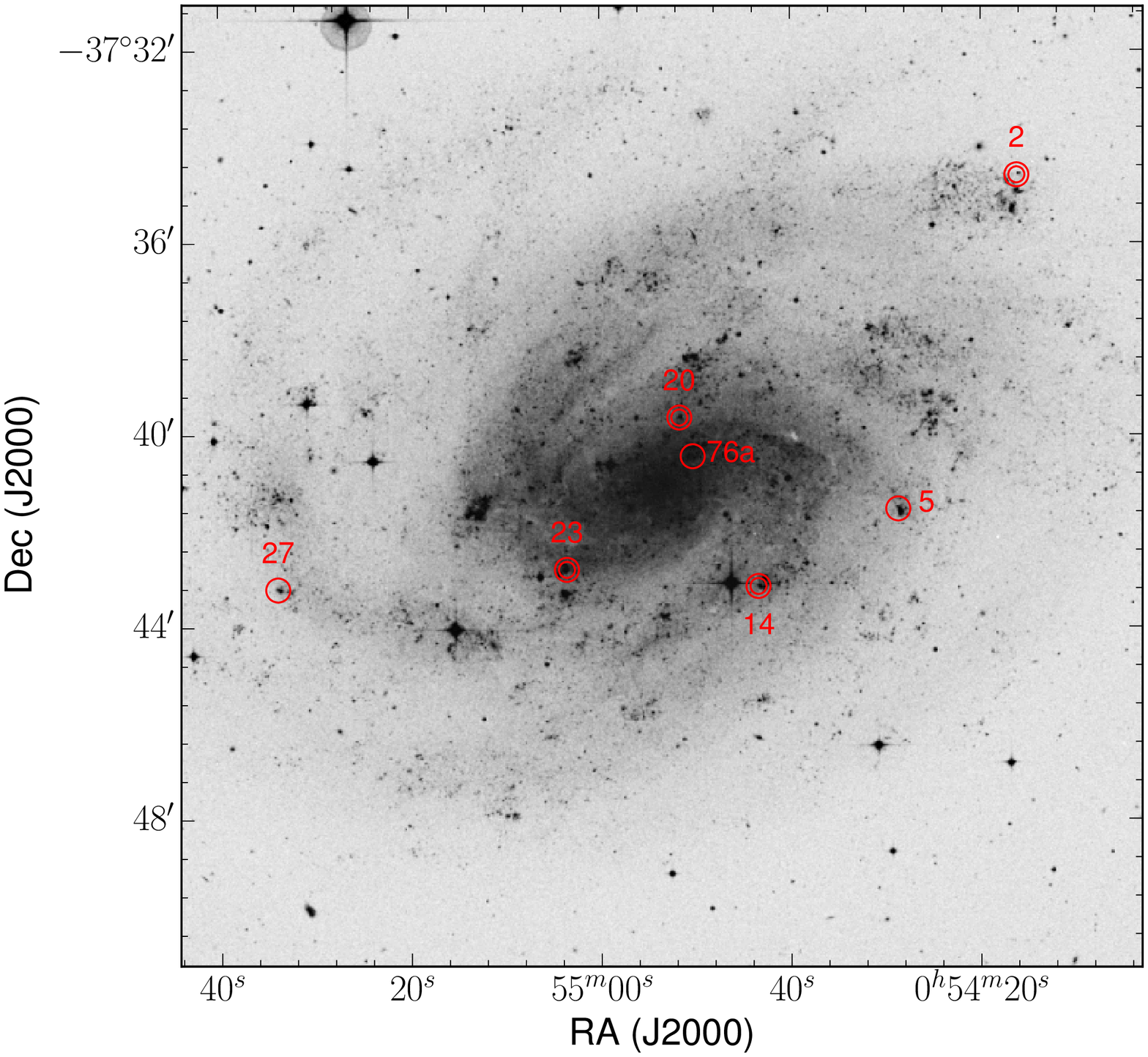}
   \includegraphics[scale=0.39]{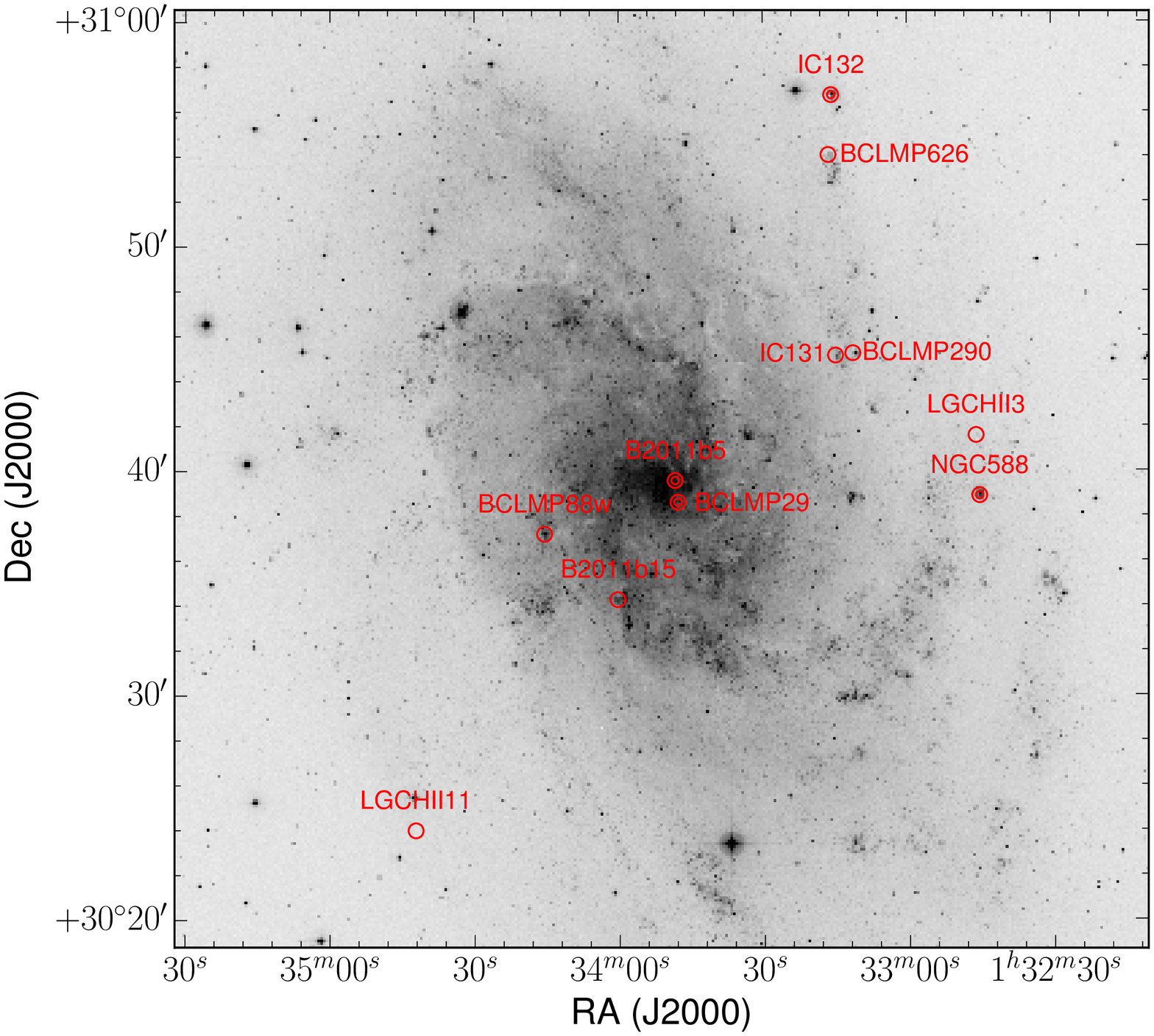}
  \caption{Distribution of the observed \hii\ regions in the disc of NGC\,300 (left panel) and M\,33 (right panel). The \hii\ region identification numbers are those used by \citet{2009ApJ...700..309B} and \citet{1988A&AS...73..407D} for NGC\,300 and by \citet{2007AA...470..865M, 2010A&A...512A..63M} and \citet{2011ApJ...730..129B} for M\,33. The double circles mark the \hii\ regions where we detected  the \cii\ 4267 \AA\ recombination line.} 
  \label{fig:hiiregions}
\end{figure*}

We selected a sample of seven \hii\ regions covering a wide range of Galactocentric distances along the disc of the NGC\,300 galaxy. Our targets were taken from \citet{2009ApJ...700..309B} and \citet{1988A&AS...73..407D}, according to their high surface brightness and their high ionization degree --high [\oiii]/\hb ratio. This was made in order to ease the detection of the faint \cii\ and \oii\ RLs in 
the objects. The distribution of the observed \hii\ regions in the disc of NGC\,300 is shown in Figure~\ref{fig:hiiregions} (left panel). 

The spectra were obtained in service time during six different nights on 2010 July and August at Cerro Paranal Observatory (Chile) using the Ultraviolet Visual Echelle Spectrograph (UVES) mounted at the Kueyen unit of the 8.2 m Very Large Telescope (VLT). We covered the spectral range 3100--10420 \AA\ using two standard configurations; DIC1(346+580) and DIC2(437+860). The wavelength intervals 5782--5824 \AA\ and 8543--8634 \AA\ were not observed due to a gap between the two CCDs used in the red arm. Additionally, there are also two small spectral gaps between 10083--10090 \AA\ and 10251--10263 \AA\ because the redmost orders did not fit completely within the CCD.  The total exposure times for each configuration and object were 1800 s for DIC1, divided in three consecutive 600 s exposures, and 4500 s for DIC2, divided in three consecutive 1500 s exposures. We also took short exposures of 30 s in DIC1 and 60 s in DIC2 to avoid saturation of the brightest lines. The slit width was set to 3\arcsec\ and the slit length was 10\arcsec\ in the red arm and 12\arcsec\ in the blue arm. This slit width was chosen to maximize the signal-to-noise ratio (S/N) and to have enough spectral resolution to separate the relevant faint lines for this study. The position angle and the common area extracted for each \hii\ region were selected to cover most of their core extension.  Spectrophotometric standard stars LTT\,377 and EG\,274 were observed for flux calibration. Table~\ref{tab:hii_data} shows the identification numbers, coordinates, deprojected Galactocentric distances -- in terms of the $R_{25}$ radius -- and the extracted area for each region of the sample of \hii\ regions observed in NGC\,300. The value  of the 25th magnitude $B$-band isophotal radius ($R_{25}$) is assumed to be 9.75 arcmin, corresponding to 5.33 kpc at the distance of 1.88 Mpc \citep{2009ApJ...700..309B}. 

The raw frames were reduced using the public ESO UVES pipeline \citep{2000Msngr.101...31B} under the \textsc{gasgano} graphic user interface, following the standard procedure of bias subtraction, order tracing, aperture extraction, background substraction, flat-fielding and wavelength calibration. The final products of the pipeline were 2D wavelength calibrated spectra. We used \textsc{iraf}\footnote{\textsc{iraf} is distributed by National Optical Astronomy Observatory, which is operated by Association of Universities for Research in Astronomy, under cooperative agreement with the National Science Foundation.} and our own \textsc{python} scripts to obtain the final one-dimensional flux calibrated spectra.

\begin{table*}
 \centering
   \caption{Sample of \hii\ regions observed in NGC\,300 and M\,33.}
    \begin{tabular}{@{}l c c c c c}
      \hline
      \multicolumn{6}{c}{NGC\,300}\\
      \hline
       & R.A.         & Decl.        & & PA            & Area\\
        ID            & (J2000)      & (J2000)      &         $R/R_{25}$  & ($\degr$)  & ($\mathrm{arcsec^2}$)\\
      \hline
        R76a         & 00 54 50.30  & -37 40 27.1  & 0.08       & 90            & $6.1 \times 3$\\
        R20         & 00 54 51.69  & -37 39 38.7  & 0.18       & 90            & $7.8 \times 3$\\
        R23         & 00 55 03.56  & -37 42 48.5  & 0.28       & 155           & $5.4 \times 3$\\ 
        R14         & 00 54 43.42  & -37 43 09.3  & 0.38       & 45            & $6.6 \times 3$\\
        R5          & 00 54 28.70  & -37 41 32.8  & 0.55       & 0             & $5.9 \times 3$\\
        R27         & 00 55 33.94  & -37 43 12.8  & 0.86       & 45            & $8.8 \times 3$\\ 
        R2          & 00 54 16.28  & -37 34 34.9  & 1.01       & 135           & $4.4 \times 3$\\
      \hline
    
      \multicolumn{6}{c}{M\,33}\\
      \hline
        B2011\,b5         & 01 33 48.4 & +30 39 35 & 0.03   &    9      & $5.9 \times 1$\\
        BCLMP\,29         & 01 33 47.8 & +30 38 37 & 0.04   &    9      & $13.8 \times 1$\\
        B2011\,b15        & 01 34 00.3 & +30 34 17 & 0.28   &   48      & $11.0 \times 1$\\ 
        BCLMP\,88w        & 01 34 15.5 & +30 37 12 & 0.34   &   48      & $11.5 \times 1$\\
        IC\,131           & 01 33 15.0 & +30 45 09 & 0.55   &   86      & $8.3 \times 1$\\
        BCLMP\,290        & 01 33 11.4 & +30 45 15 & 0.60   &   86      & $15.8 \times 1$\\ 
        NGC\,588          & 01 32 45.2 & +30 38 54 & 0.79   &    0      & $14.8 \times 1$\\
        BCLMP\,626        & 01 33 16.4 & +30 54 05 & 0.83   &   -3      & $5.3 \times 1$\\ 
        LGC\,\hii3        & 01 32 45.9 & +30 41 35 & 0.83   &    0      & $4.8 \times 1$\\
        IC\,132           & 01 33 15.8 & +30 56 45 & 0.93   &    -3     & $9.4 \times 1$\\
        LGC\,\hii11       & 01 34 42.2 & +30 24 00 & 1.03   &    90     & $3.9 \times 1$\\
      \hline

    \end{tabular}
    \label{tab:hii_data}   
\end{table*}

The observations of M\,33 were made on 2013 November during nine different nights in service time with the Optical System for Imaging and low-Intermediate-Resolution Integrated Spectroscopy (OSIRIS) instrument \citep{2000SPIE.4008..623C, 2003SPIE.4841.1739C} on the 10.4 m Gran Telescopio de Canarias (GTC). We selected a sample of eleven \hii\ regions distributed across the disk of the galaxy (see Fig. \ref{fig:hiiregions}, right panel). The targets were selected following the same criteria than for NGC\,300 based in the work of \citet{2007AA...470..865M, 2010A&A...512A..63M} and \citet{2011ApJ...730..129B}. We covered the spectral range 3600--7600 \AA\ using three grisms: R2500U, R2500V and R2500R. The total exposure times for each configuration and object were 2610 s for R2500U, divided in three consecutive 870 s exposures, 3480 s for R2500V, divided in four consecutive 870 s exposures, and 1650 s for R2500R, divided in three consecutive 550 s exposures. We took also short exposures of 60 s in the three grisms to avoid saturation of the brightest lines. The slit width was set to 1\arcsec\ and the slit length was 7.4\arcmin. The position angles and the common areas extracted for each \hii\ region were selected to cover most of their core extension. For the flux calibrations two spectrophotometric standard stars: Hiltner 600 and G191-B2B were observed. In Table \ref{tab:hii_data} we present the information of the sample of \hii\ regions observed in M\,33; the identification names, coordinates, deprojected Galactocentric distances -- in terms of the $R_{25}$ radius -- the position angle and the extracted area. We adopted the value of the isophotal radius ($R_{25}$) of 28\arcmin\ \citep{1976RC2...C......0D}.

The raw frames were reduced using \textsc{iraf} following the standard procedure: bias subtraction, flat-fielding and wavelength calibration one-dimensional spectra, extraction and flux calibration.

\section{Line fluxes, identification and reddening correction}\label{sec:iden}

\begin{figure*}
  \includegraphics{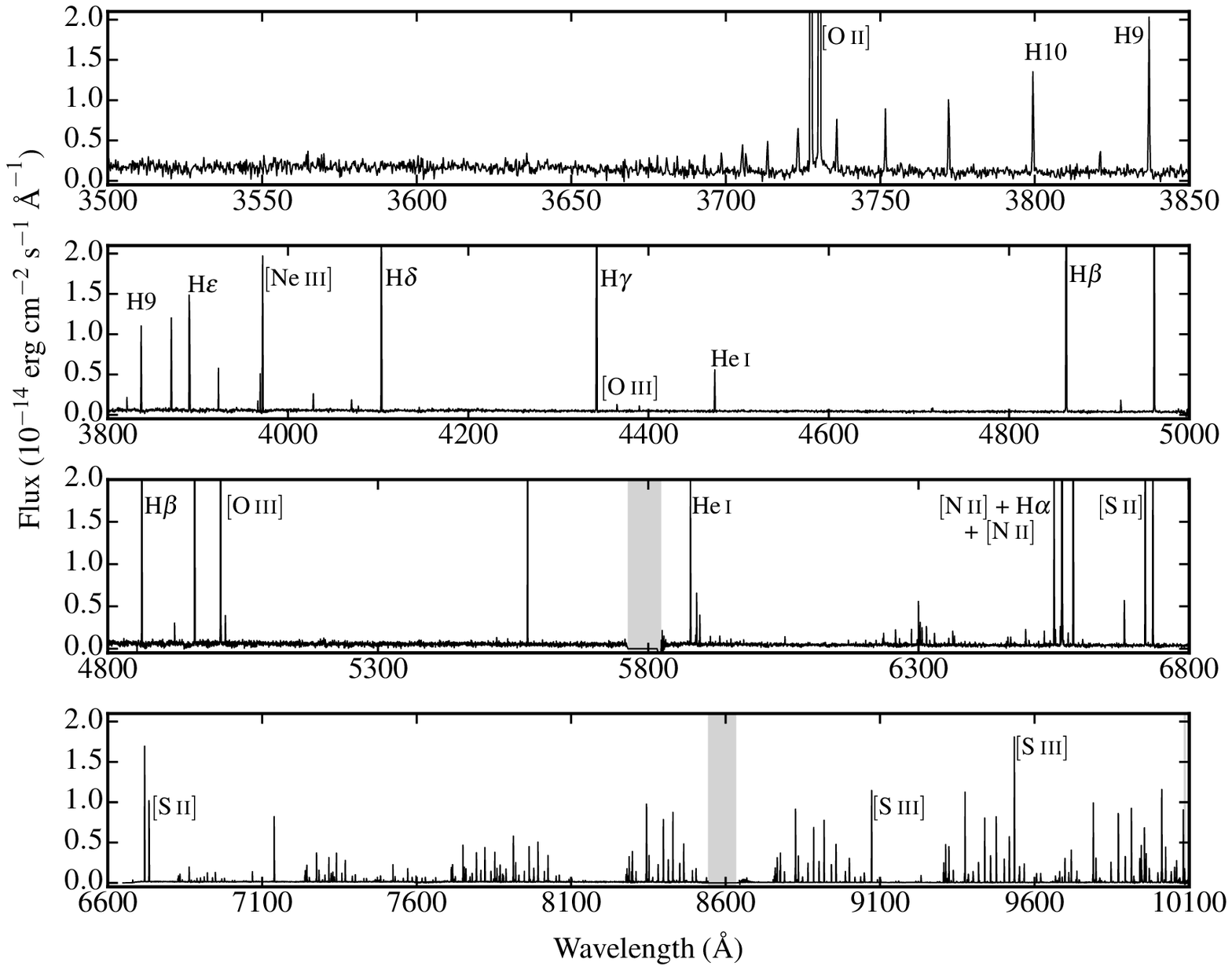}
  \caption{Flux-calibrated VLT UVES spectrum of the \hii\ region R14 of NGC\,300. The shaded wavelength ranges represent the gaps described in Section \ref{sec:obs}. Sky emission has not 
  been removed.} 
  \label{fig:spec}
\end{figure*}

\begin{figure*}
  \includegraphics{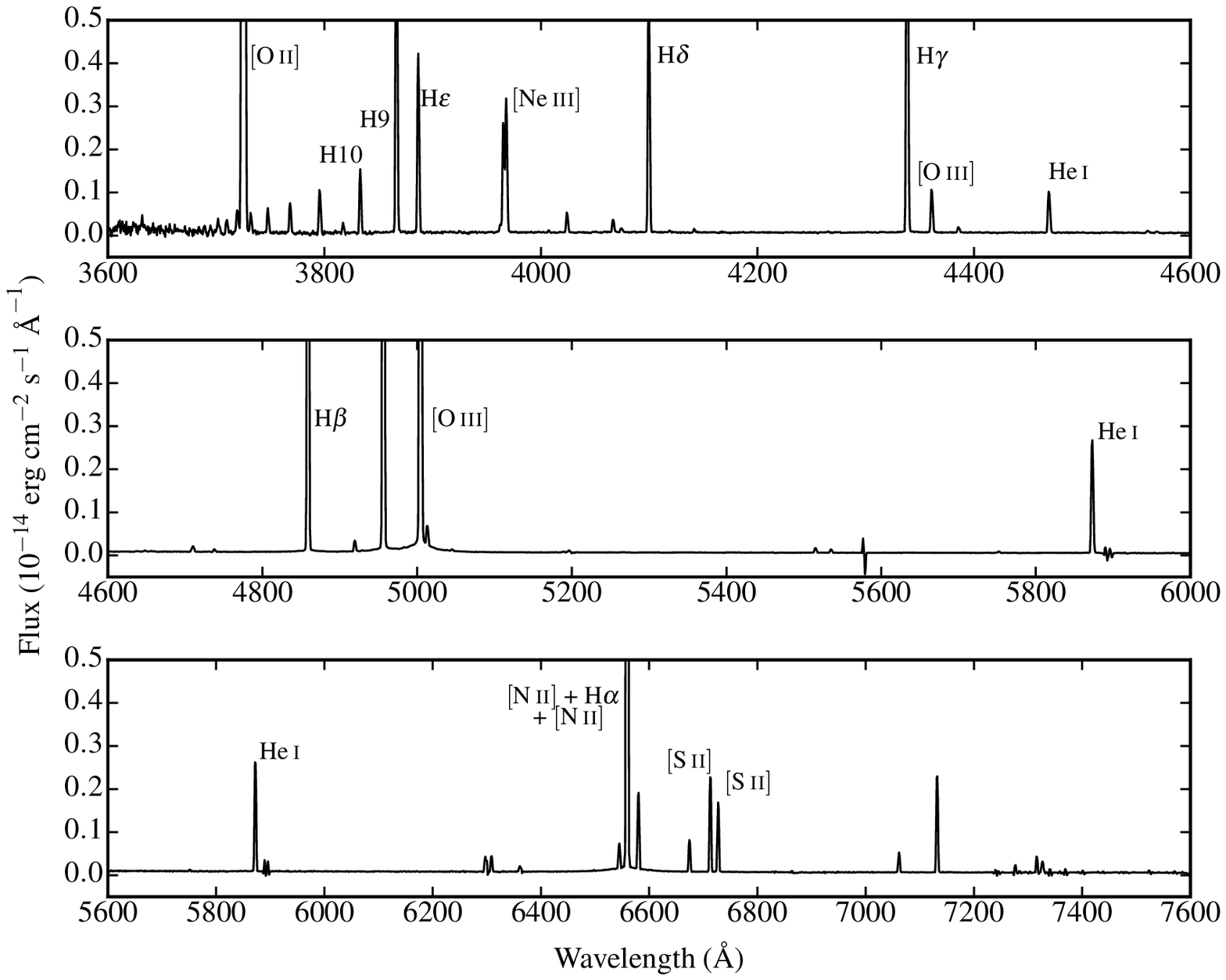}
  \caption{Flux-calibrated GTC OSIRIS spectrum of the \hii\ region NGC588 of M\,33. Sky emission has been removed.} 
  \label{fig:spec_m33}
\end{figure*}

We used the {\sc splot} routine of {\sc iraf} to measure the line fluxes. We integrated all the flux in the line between two given limits and over a local continuum estimated by eye. In case of line blending we fitted a double Gaussian profile to measure the individual line fluxes.

All the line fluxes were normalized to the H$\beta$ flux. For NGC\,300, the flux normalization was made directly in DIC2 437 and DIC1 580 configurations because H$\beta$ is observed in both of them. For DIC1 346 all the line fluxes were normalized to the H9 line and then we used the H9/H$\beta$ ratio observed in DIC2 437 to re-scale to H$\beta$. In  DIC2 860 configuration there are no bright {\hi} recombination lines in common with the DIC1 580 spectrum. Therefore, we normalized the line fluxes to \fsii\ 6730 \AA\ and then we used the \fsii 6730 \AA/\hb\ ratio observed in DIC1 580 configuration to re-scale to H$\beta$. The line fluxes of lines in the reddest part of the spectrum of DIC2 860 were normalized with respect to the bright P9 \hi\  Paschen line and then re-scaled to \hb\ using the theoretical flux ratio for $T_\mathrm{e}=10000$ K and $n_\mathrm{e}=100\ \mathrm{cm^{-3}}$  \citep{1995MNRAS.272...41S} and applying the extinction law by \citet{1989ApJ...345..245C} (see below). 

The flux normalization in M33 was made directly in R2500V grism because H$\beta$ is observed in this configuration. For R2500U grism, the line fluxes were normalized to the \hei\ 4471 \AA\ line and then we used the \hei\ 4471/H$\beta$ ratio observed in R2500V to re-scale to H$\beta$. The procedure followed in R2500R was the same that for R2500U. In this case, we normalized the line fluxes to \hei\ 5875 \AA, and then we re-scale to H$\beta$ with the \hei\ 5875/H$\beta$ ratio observed in R2500V.

The observed flux ratios must be corrected for interstellar reddening. We  determined the reddening coefficient, $c$({\hb}), following the equation: 

\begin{equation}
c(\mathrm{H}\beta) = \frac{1}{f(\lambda)}\log \left[\frac{\frac{I(\lambda)}{I(\mathrm{H}\beta)}  \left(1 + \frac{W_{\mathrm{abs}}}{W_{\mathrm{H}\beta}}\right)}{\frac{F(\lambda)}{F(\mathrm{H}\beta)} \left(1 + \frac{W_{\mathrm{abs}}}{W_{\lambda}}\right)}\right] \label{eq: c(Hbeta)}
\end{equation}
given by \citet{2006A&A...449..997L}  which considers that the \hi\ lines are affected by underlying stellar absorption. $F(\lambda)$/$F$({\hb}) is the observed flux ratio with respect to \hb, $I(\lambda)$/$I$({\hb}) is the theoretical ratio, $W_{\mathrm{abs}}$,  $W_{\lambda}$ and $W_{\rm H\beta}$ are the equivalent widths of the underlying stellar absorption,  the considered Balmer line and H$\beta$, respectively, and $f(\lambda)$ is the reddening function. To compute the $c$({\hb}) with this equation we used the brightest Balmer lines (H$\alpha$, H$\gamma$ and H$\delta$ with respect to {\hb}) and we assumed that the $W_{\mathrm{abs}}$ is the same for all these lines. The theoretical line ratios were obtained from \citet{1995MNRAS.272...41S} for case B assuming the physical conditions computed by \citet{2009ApJ...700..309B} for NGC\,300 \hii\ regions and by \citet{2007AA...470..865M, 2010A&A...512A..63M} and \citet{2011ApJ...730..129B} for M\,33 \hii\ regions. We used the reddening function, $f(\lambda)$, by \citet{1989ApJ...345..245C} with $R_V = 3.1$. We performed an iterative process to derive $c$({\hb}) with different $W_{\mathrm{abs}}$ and chosing the $c$({\hb}) that best fits the observed/theoretical line ratios. Once we calculated the reddening coefficient, the observed flux ratios were corrected. The value of the adopted $c$({\hb}), as well as of the adopted $W_{\mathrm{abs}}$, are included at the end of Tables~\ref{tab:data_1} and  \ref{tab:data_2} for \hii\ regions in NGC\,300 and Tables~\ref{tab:data_1_m33}, \ref{tab:data_2_m33} and \ref{tab:data_3_m33} for \hii\ regions in M\,33.

Tables~\ref{tab:data_1}, \ref{tab:data_2}, \ref{tab:data_1_m33}, \ref{tab:data_2_m33} and \ref{tab:data_3_m33} show all the emission-line ratios measured for each \hii\ region. The regions are ordered from the most internal region to the more external one in each galaxy. Column 1 gives the laboratory wavelength, columns 2 and 3 indicate the ion and multiplet, column 4 gives the $f(\lambda)$ value. The observed wavelength and the reddening corrected flux (in units of $I(\mathrm{H\beta})=100$) of the lines and their associated uncertainty for all the {\hii} regions are given in subsequent pairs of columns after column 5. To compute flux errors we added quadratically the uncertainties in the measurement of the line intensity and the estimated flux calibration error ($\sim$5\%). Then we propagated the error in the reddening coefficients to obtain the final uncertainties. The identification and laboratory wavelength of the lines were obtained following previous works by \citet{2004ApJS..153..501G} and  \citet{2004MNRAS.355..229E}. In Figure~\ref{fig:spec} we show the flux-calibrated VLT UVES spectra of the \hii\ region R14 of NGC\,300. The  shaded narrow wavelength ranges areas with zero continuum flux correspond to the gaps in the ccd described in Section \ref{sec:obs}. In Figure \ref{fig:spec_m33} we show the flux-calibrated GTC OSIRIS spectrum of the \hii\ region NGC\,588 of M\,33. The most prominent features in the spectrum have been labelled. 

\onecolumn
\begin{landscape}
\begin{longtable}{c c c c c c c c c c c c}
         \caption{Line identifications and reddening corrected line fluxes with respect to $\mathrm{I}(H\beta) = 100$ for the \hii\ regions R76a, R20, R23 and R14 of NGC\,300.}\\
          \label{tab:data_1}\\
             \hline
                             &      &      &              &\multicolumn{2}{c}{R76a}                 &\multicolumn{2}{c}{R20}                   &  \multicolumn{2}{c}{R23}                 & \multicolumn{2}{c}{R14}       \\                                                        
             $\lambda_0$     &      &      &              &$\lambda_{\mathrm{obs}}$ &               &$\lambda_{\mathrm{obs}}$ &                &$\lambda_{\mathrm{obs}}$ &                &$\lambda_{\mathrm{obs}}$ &  \\
                (\AA)        & Ion  &  ID  & $f(\lambda)$ & (\AA)                   &$I$($\lambda$) &  (\AA)                  & $I$($\lambda$) &        (\AA)            & $I$($\lambda$) &   (\AA)                 &$I$($\lambda$)  \\
             \hline
    \endfirsthead
          \caption{continued.}\\
           \hline
                             &      &      &              &\multicolumn{2}{c}{R76a}                 &\multicolumn{2}{c}{R20}                   &  \multicolumn{2}{c}{R23}                 & \multicolumn{2}{c}{R14}       \\
             $\lambda_0$     &      &      &              &$\lambda_{\mathrm{obs}}$ &               &$\lambda_{\mathrm{obs}}$ &                &$\lambda_{\mathrm{obs}}$ &                &$\lambda_{\mathrm{obs}}$ &  \\
                (\AA)        & Ion  &  ID  & $f(\lambda)$ & (\AA)                   &$I$($\lambda$) &  (\AA)                  & $I$($\lambda$) &        (\AA)            & $I$($\lambda$) &   (\AA)                 &$I$($\lambda$)  \\
            \hline
    \endhead
            \hline
    \endfoot
    \endlastfoot
            3187.84       & \hei   &   3   &0.469  & --     & --            &3188.98&6.11 $\pm$ 1.08&3188.49& 4.3 $\pm$ 0.6 &3189.06&4.3 $\pm$ 0.6 \\
            3613.64       & \hei   &   6   &0.344  & --     &--             & --    &  --           &3614.46& 0.39:         &3615.11&0.58:    \\     
            3634.25       & \hei   &   28  &0.340  & --     &--             & --    &  --           &3635.19& 0.28:         &3635.68&0.44:  \\     
            3663.4        & \hi    &  H29  &0.335  & --      &--             & --    &  --           &3664.20& 0.32:         &--    &--              \\     
            3664.68       & \hi    &  H28  &0.334  & --     &--             & --    &  --           &3665.53& 0.32:         &--    &--              \\     
            3666.1        & \hi    &  H27  &0.334  & --     &--             & --    &  --           &3667.04& 0.31:         &--    &--              \\     
            3667.68       & \hi    &  H26  &0.334  & --     &--             & --    &  --           &3668.54& 0.37:         &--    &--              \\  
            3669.47       & \hi    &  H25  &0.334  & --     & --            &3670.89 &0.49:         &3670.29& 0.40:         &3670.97&0.44:          \\
\end{longtable}
\end{landscape}  
\twocolumn

\onecolumn
\begin{landscape}
  \begin{longtable}{cccccccccc}
      \caption{Line identifications and reddening corrected line fluxes with respect to I($H\beta$) = 100 for the \hii\ regions R5, R27, and R2 of NGC\,300.}\\
      \label{tab:data_2}\\
         \hline
                     &     &    &              &\multicolumn{2}{c}{R5}                  & \multicolumn{2}{c}{R27}                   &  \multicolumn{2}{c}{R2}                \\
         $\lambda_0$ &     &    &              &$\lambda_{\mathrm{obs}}$ &              &$\lambda_{\mathrm{obs}}$   &               &$\lambda_{\mathrm{obs}}$ &               \\
         (\AA)       &Ion  & ID &$f(\lambda)$  &  (\AA)                  &$I$($\lambda$)&  (\AA)                    &$I$($\lambda$) & (\AA)                   &$I$($\lambda$) \\
         \hline
      \endfirsthead
         \caption{continued.}\\
         \hline
                     &     &    &              &\multicolumn{2}{c}{R5}                  & \multicolumn{2}{c}{R27}                   &  \multicolumn{2}{c}{R2}                \\
         $\lambda_0$ &     &    &              &$\lambda_{\mathrm{obs}}$ &              &$\lambda_{\mathrm{obs}}$   &               &$\lambda_{\mathrm{obs}}$ &               \\
         (\AA)       &Ion  & ID &$f(\lambda)$  &  (\AA)                  &$I$($\lambda$)&  (\AA)                    &$I$($\lambda$) & (\AA)                   &$I$($\lambda$) \\
         \hline
      \endhead
         \hline
      \endfoot
      \endlastfoot
3187.84 & \hei   &   3   &0.469 &--    &--              &--    &--               &3189.26&8.5 $\pm$ 1.3\\
3671.48 & \hi   &  H24  &0.333  &--    &--              &--    &--               &3673.23&0.71:        \\
3673.76 & \hi    &  H23  &0.333 &--    &--              &--    &--               &3675.56&0.65:          \\
3676.37 & \hi    &  H22  &0.332 &--    &--              &--    &--               &3678.03&0.99:          \\
3679.36 & \hi    &  H21  &0.332 &3681.19&0.98:          &--    &--               &3681.20&0.81:          \\
3682.81 & \hi    &  H20  &0.331 &3684.73&1.1:           &3683.26&1.3:            &3684.58&0.75:          \\
3686.83 & \hi    &  H19  &0.330 &--    &--              &3687.79&1.4:            &3688.47&1.2:          \\
3691.56 & \hi   &  H18  &0.329  &3693.46&1.1:           &--     &--              &3693.34&1.6:          \\
\end{longtable}
\end{landscape}  
\twocolumn

\onecolumn
\begin{landscape}
  \begin{longtable}{c c c c c c c c c c c c}
         \caption{Line identifications and reddening corrected line fluxes with respect to $\mathrm{I}(H\beta) = 100$ for the \hii\ regions B2011\,b5, BCLMP\,29, B2011\,b15 and BCLMP\,88w of M\,33.}\\
          \label{tab:data_1_m33}\\
             \hline
                             &      &      &              &\multicolumn{2}{c}{B2011\,b5}            &\multicolumn{2}{c}{BCLMP\,29}                   &  \multicolumn{2}{c}{B2011\,b15}               & \multicolumn{2}{c}{BCLMP\,88w}       \\                                                        
             $\lambda_0$     &      &      &              &$\lambda_{\mathrm{obs}}$ &               &$\lambda_{\mathrm{obs}}$ &                &$\lambda_{\mathrm{obs}}$ &                &$\lambda_{\mathrm{obs}}$ &  \\
                (\AA)        & Ion  &  ID  & $f(\lambda)$ & (\AA)                   &$I$($\lambda$) &  (\AA)                  & $I$($\lambda$) &        (\AA)            & $I$($\lambda$) &   (\AA)                 &$I$($\lambda$)  \\
             \hline
    \endfirsthead
          \caption{continued.}\\
           \hline
                             &      &      &              &\multicolumn{2}{c}{B2011\,b5}            &\multicolumn{2}{c}{BCLMP\,29}                   &  \multicolumn{2}{c}{B2011\,b15}               & \multicolumn{2}{c}{BCLMP\,88w}       \\
             $\lambda_0$     &      &      &              &$\lambda_{\mathrm{obs}}$ &               &$\lambda_{\mathrm{obs}}$ &                &$\lambda_{\mathrm{obs}}$ &                &$\lambda_{\mathrm{obs}}$ &  \\
                (\AA)        & Ion  &  ID  & $f(\lambda)$ & (\AA)                   &$I$($\lambda$) &  (\AA)                  & $I$($\lambda$) &        (\AA)            & $I$($\lambda$) &   (\AA)                 &$I$($\lambda$)  \\
            \hline
    \endhead
            \hline
    \endfoot
    \endlastfoot
            3691.56       & \hi    &  H18  & 0.329 &  --    &       --                 &3689.95  &1.28 $\pm$ 0.25            &    --  &      --                 &  --    &     --        \\ 
            3697.15        & \hi    &  H17  & 0.328 &  --    &       --                 &3689.95  &1.12 $\pm$ 0.23            &    --  &      --                 &  --    &     --        \\ 
            3703.86       & \hi    &  H16  &0.326  &   --   &       --                 &3702.35  &2.34 $\pm$ 0.35            &   --   &      --                 &3702.49 &1.88 $\pm$ 0.55\\
            3705.04       & \hei   &  25   &       &      &                        &    *    &      *                    &      &                       &    *   &        *       \\        
            3711.97       & \hi    &  H15  &0.325  &3709.89 &1.96 $\pm$ 0.36           &3710.02  &1.23 $\pm$ 0.24            &   --   &      --                 &3710.00 &1.16:          \\
            3721.83       & [\siii]&  2F   &0.323  &3720.17 &2.13 $\pm$ 0.38           &3720.00  &1.87 $\pm$ 0.30            &  --    &      --                 &3719.99 &1.72 $\pm$ 0.51\\
            3721.94       & \hi    &  H14  &        &   *    &        *                 &   *     &        *                  &     &                       &   *    &  *             \\
            3726.03       & [\oii] &  1F   &0.322  &3724.01 & 68  $\pm$  10            &3724.07  &  76 $\pm$  6              &3723.39 &110  $\pm$  12           &3724.00 &   41 $\pm$ 9  \\
\end{longtable}
\end{landscape}  
\twocolumn


\onecolumn
\begin{landscape}
  \begin{longtable}{c c c c c c c c c c c c}
         \caption{Line identifications and reddening corrected line fluxes with respect to $\mathrm{I}(H\beta) = 100$ for the \hii\ regions IC\,131, BCLMP\,290, NGC\,588 and BCLMP\,626 of M\,33.}\\
          \label{tab:data_2_m33}\\
             \hline
                             &      &      &              &\multicolumn{2}{c}{IC\,131}                 &\multicolumn{2}{c}{BCLMP\,290}                   &  \multicolumn{2}{c}{NGC\,588}                 & \multicolumn{2}{c}{BCLMP\,626}       \\                                                        
             $\lambda_0$     &      &      &              &$\lambda_{\mathrm{obs}}$ &               &$\lambda_{\mathrm{obs}}$ &                &$\lambda_{\mathrm{obs}}$ &                &$\lambda_{\mathrm{obs}}$ &  \\
                (\AA)        & Ion  &  ID  & $f(\lambda)$ & (\AA)                   &$I$($\lambda$) &  (\AA)                  & $I$($\lambda$) &        (\AA)            & $I$($\lambda$) &   (\AA)                 &$I$($\lambda$)  \\
             \hline
    \endfirsthead
          \caption{continued.}\\
           \hline
                             &      &      &              &\multicolumn{2}{c}{IC\,131}                 &\multicolumn{2}{c}{BCLMP\,290}                   &  \multicolumn{2}{c}{NGC\,588}                 & \multicolumn{2}{c}{BCLMP\,626}       \\
             $\lambda_0$     &      &      &              &$\lambda_{\mathrm{obs}}$ &               &$\lambda_{\mathrm{obs}}$ &                &$\lambda_{\mathrm{obs}}$ &                &$\lambda_{\mathrm{obs}}$ &  \\
                (\AA)        & Ion  &  ID  & $f(\lambda)$ & (\AA)                   &$I$($\lambda$) &  (\AA)                  & $I$($\lambda$) &        (\AA)            & $I$($\lambda$) &   (\AA)                 &$I$($\lambda$)  \\
            \hline
    \endhead
            \hline
    \endfoot
    \endlastfoot
            3686.83       & \hi    &  H19  &0.330  &  --    &      --                 &  --    &      --                 &3684.17 &0.28 $\pm$ 0.05          &   --    &  --\\
            3691.56       & \hi    &  H18  &0.329  &  --    &      --                 &3689.18 &0.98 $\pm$ 0.12          &3689.25 &0.58 $\pm$ 0.09          &   --    &  --\\
            3697.15       & \hi    &  H17  &0.328  &  --    &      --                 &3694.44 &1.01 $\pm$ 0.12          &3694.86 &0.52 $\pm$ 0.08          &   --    &  --\\
            3703.86       & \hi    &  H16  &0.327  &  --    &      --                 &3701.37 &1.42 $\pm$ 0.15          &3701.72 &1.30 $\pm$ 0.14          &   --    &  --\\
            3705.04       & \hei   &  25   &  &      &                       &  *     &     *                   &    *   &  *                      &       &  \\
            3711.97       & \hi    &  H15  &0.325  &   --   &       --                &3709.11 &0.98 $\pm$ 0.12          &3709.82 &1.24 $\pm$ 0.16          &   --    &  --\\
            3721.83       & [\siii]&  2F   &0.323  &3719.12 &2.72 $\pm$ 0.52          &3719.27 &1.67 $\pm$ 0.17          &3719.56 &2.37 $\pm$ 0.28          &   --    &  --\\
            3721.94       & \hi    &  H14  &        &   *    &       *                 &   *    &       *                 &   *    &     *                   &       &  \\
\end{longtable}
\end{landscape}  
\twocolumn

\onecolumn
\begin{landscape}
  \begin{longtable}{c c c c c c c c c c}
         \caption{Line identifications and reddening corrected line fluxes with respect to $\mathrm{I}(H\beta) = 100$ for the \hii\ regions LGC\,\hii3, IC\,132 and LGC\,\hii11  of M\,33.}\\
          \label{tab:data_3_m33}\\
             \hline
                             &      &      &              &\multicolumn{2}{c}{LGC\,\hii3}                 &\multicolumn{2}{c}{IC\,132}                   &  \multicolumn{2}{c}{LGC\,\hii11}   \\                                                        
             $\lambda_0$     &      &      &              &$\lambda_{\mathrm{obs}}$ &               &$\lambda_{\mathrm{obs}}$ &                &$\lambda_{\mathrm{obs}}$ &                   \\
                (\AA)        & Ion  &  ID  & $f(\lambda)$ & (\AA)                   &$I$($\lambda$) &  (\AA)                  & $I$($\lambda$) &        (\AA)            & $I$($\lambda$)     \\
             \hline
    \endfirsthead
          \caption{continued.}\\
           \hline
                             &      &      &              &\multicolumn{2}{c}{LGC\,\hii3}                 &\multicolumn{2}{c}{IC\,132}                   &  \multicolumn{2}{c}{LGC\,\hii11}                 \\
             $\lambda_0$     &      &      &              &$\lambda_{\mathrm{obs}}$ &               &$\lambda_{\mathrm{obs}}$ &                &$\lambda_{\mathrm{obs}}$ &                         \\
                (\AA)        & Ion  &  ID  & $f(\lambda)$ & (\AA)                   &$I$($\lambda$) &  (\AA)                  & $I$($\lambda$) &        (\AA)            & $I$($\lambda$)      \\
            \hline
    \endhead
            \hline
    \endfoot
    \endlastfoot
            3703.86       & \hi    &  H16  &0.327  &3701.99 &1.87 $\pm$ 0.39&3701.64 &2.12 $\pm$ 0.31          &3701.97 &3.04 $\pm$ 0.93 \\
            3705.04       & \hei   &  25   &        &  *     &      *        &  *     &       *                 &   *    &         *  \\
            3711.97       & \hi    &  H15  &0.325  &3709.68 &1.35 $\pm$ 0.36&3709.56 &1.20 $\pm$ 0.18          &  --    &     --      \\
            3721.83       & [\siii]&  2F   &0.323  &3719.53 &2.54 $\pm$ 0.44&3719.29 &2.18 $\pm$ 0.32          &3719.69 &2.86 $\pm$ 0.90\\
            3721.94       & \hi    &  H14  &        &   *    &     *         &   *    &       *                 &    *   &     *         \\
            3726.03       & [\oii] &  1F   &0.322  &3723.70 &  72 $\pm$ 6   &3723.66 &  26 $\pm$ 3             &3723.83 & 115 $\pm$ 20\\
            3728.82       & [\oii] &  1F   &0.328  &3726.49 & 111 $\pm$ 9   &3726.42 &  35 $\pm$ 4             &3726.61 & 146 $\pm$ 26\\
            3734.37       & \hi    &  H13  &0.320  &3731.91 &1.55 $\pm$ 0.37&3731.82 &2.32 $\pm$ 0.33          &3732.27 &2.58 $\pm$ 0.86\\
\end{longtable}
\end{landscape}  
\twocolumn

\section{Physical conditions of the ionized gas} \label{sec:physicalcondition}

The physical conditions of the ionized gas, electron density and temperature, $n_\mathrm{e}$ and $T_\mathrm{e}$, were derived using PyNeb \citep{2015A&A...573A..42L} and the atomic data set shown in Table~\ref{tab:atomicData}.

\begin{table}
     \caption{Sources of Atomic Data.}
     \label{tab:atomicData}
     \begin{tabular}{@{}l cc}
           \hline
           Ion & Transition Probabilities & Collision Strengths  \\
           \hline
           \foii  & Froese Fischer     & Tayal (2007) \\
                  & \& Tachiev (2004)  &  \\
           \foiii & Froese Fischer     &  Aggarwal \& Keenan (1999)   \\
                  & \& Tachiev (2004)  &  \\
           \fnii  & Froese Fischer     & Tayal (2011) \\ 
                  & \& Tachiev (2004)  &  \\
           \fneiii&Galav\'\i s, Mendoza    & McLaughlin \& Bell (2000) \\
                  &\& Zeippen (1997)   &          \\
           \fariii& Mendoza (1983)     &Galav\'\i s, Mendoza\\
                  & Kaufman \& Sugar (1986)   &  \& Zeippen (1995) \\
           \fariv & Mendoza \& Zeippen(1982) & Ramsbottom \& Bell (1997)\\
           \fsii  &Podobedova, Kelleher & Ramsbottom, Bell, \\ 
                  &  \& Wiese (2009)    & \& Stafford (1996)\\
           \fsiii &Podobedova, Kelleher & Tayal \& Gupta (1999)  \\
                  & \& Wiese (2009)     & \\
           \fcliii& Mendoza (1983)   & Butler \& Zeippen (1989) \\
                  &Kaufman \& Sugar (1986) &        \\
           \ffeiii& Quinet (1996)       & Quinet (1996) \\
           \hline
 \end{tabular}
 
 \medskip 
\end{table}

In Table~\ref{tab:physical_conditions} we show the physical conditions of the ionized gas for each \hii\ region of NGC\,300 and M\,33. The electron density was computed through \fsii\ 6731/6716, \foii\ 3726/3729 and \fcliii\ 5538/5518 line ratios. In most cases, the $n_\mathrm{e}$ values we obtain using \fsii\ and \foii\ diagnostics are at the low density limit (below 100 cm$^{-3}$). We were able to derive $n_\mathrm{e}$ from \fcliii\ in several objects, but with large uncertainties. We then assumed a canonical density of $\sim$100 cm$^{-3}$ for all the objects in our sample with $n_\mathrm{e}$ below 100 cm$^{-3}$. To compute $T_\mathrm{e}$ we  used several diagnostic ratios when available. These diagnostics are \foiii\ 4363/5007, \fsiii\ 6312/9069, \fnii\ 5755/6584, \foii\ 7320+30/3726+29 and \fsii\ 4068+76/6717+31.

\begin{table*}
   \begin{scriptsize}
   \caption{Physical Conditions of \hii\ regions in NGC\,300 and M\,33 .}
   \label{tab:physical_conditions}
   \begin{tabular}{@{}l c c c c c c c c c c}
          \hline
                                     &\multicolumn{8}{c}{NGC\,300}\\
          
          Parameter                   & Lines         & R76a              & R20               & R23               & R14               & R5                & R27                  & R2 \\
          \hline
          $n_\mathrm{e}$ (cm$^{-3}$) & \fsii          &  --               & $<100$             &  $<100$            &  $<100$            &  --               & --                & $<100$ \\
                                     & \foii          &  $<100$            & $<100$             & --                & --                &  $<100$            &  $<100$            &  $<100$ \\
                                     & \fcliii        &  --               & 370:              & --                & --                & --                & --                & -- \\
          $T_\mathrm{e}$ (K)         & \foiii         &  8600:            & 7900 $\pm$ 400    & 7800 $\pm$ 400    &8200 $\pm$ 350     & 9500 $\pm$ 900    &  9700:            & 11500 $\pm$ 500\\
                                     & \fsiii         &  --               & 8500  $\pm$ 600   & 8100  $\pm$ 500   &9250  $\pm$ 500    & 11100  $\pm$ 1400 &  --               & 12500  $\pm$ 1600 \\
                                     &{\bf $T$(high)} &  \textbf{8600:}   &\bf{8100 $\pm$ 250}&\bf{7900 $\pm$ 300}&\bf{8500 $\pm$ 300}&\bf{9950 $\pm$ 500}&\bf{9700:}         & \bf{11600 $\pm$ 500}\\
                                     & \fnii          &  8300:            & 8400:             & 8750 $\pm$ 1300   & 8500 $\pm$ 1200   & --                & --                & --\\
                                     & \foii          & 7850 $\pm$ 250    & 9050 $\pm$  300   &  9700 $\pm$ 350   & 10000 $\pm$ 300   & 9250 $\pm$ 200    &10400 $\pm$ 350    & 10800  $\pm$ 300\\
                                     & \fsii          &--                 &  7600 $\pm$  700  &  7400 $\pm$ 550   & 7200 $\pm$  400   & 7750 $\pm$ 800    &--                 & 8600  $\pm$ 950\\
                                     & {\bf $T$(low)} &\bf{7850 $\pm$ 250}&\bf{9000 $\pm$ 350}&\bf{9600 $\pm$ 350}&\bf{9950 $\pm$ 300}&\bf{9250 $\pm$ 200}&\bf{10400 $\pm$ 250}& \bf{10800  $\pm$ 300}\\
          \hline
                                     &\multicolumn{8}{c}{M\,33}\\
           
           Parameter                  & Lines          &B2011\,b5              &BCLMP\,29            &B2011\,b15             &BCLMP\,88w           &IC\,131             &BCLMP\,290           &             \\
          \hline                                                                                                                                                                       
          $n_\mathrm{e}$ (cm$^{-3}$)  & \fsii          &  100 $\pm$ 300      & $<100$              &  $<100$             &  $<100$             &  $<100$            &  $<100$             &             \\
                                      & \foii          &  200 $\pm$ 300      & $<100$              &  $<100$             &  $<100$             &  $<100$            &  $<100$             &             \\
                                      & \fcliii        &  200:               &    100:             & --                  &   400:              &  600:              & --                  &             \\
          $T_\mathrm{e}$ (K)     &    \foiii$^{\rm a}$ & {\bf 7550 $\pm$ 300} &{\bf 8000 $\pm$ 400} &{\bf 8000 $\pm$ 900$^{\rm c}$} &{\bf 8000 $\pm$ 500} &{\bf 8900 $\pm$ 250}&{\bf 8300 $\pm$ 250} &             \\
                                      & \fnii$^{\rm b}$          & {\bf 8500 $\pm$ 400} &{\bf 7600 $\pm$ 200} &{\bf 8700 $\pm$ 900} &{\bf 8200 $\pm$ 1000}&{\bf 9000 $\pm$ 900}&{\bf 9500 $\pm$ 400} &             \\
                                      & \foii          & 10600 $\pm$ 650     & 8200 $\pm$ 300      & 8450 $\pm$ 350      & 10000 $\pm$ 1100    & 16800 $\pm$ 2000   & 10600 $\pm$ 400     &             \\                 
                                      & \fsii          &      --             & 6000 $\pm$ 600      &       --            &         --          &        --          &       --            &             \\
          \hline                                                                                                                                                                                        
                                      &                & NGC\,588            &BCLMP\,626             &LGC\,\hii3           &IC\,132                &LGC\,\hii11          &                    &             \\
          \hline                                   
          $n_\mathrm{e}$ (cm$^{-3}$)  & \fsii          & $<100$              & $<100$              & $<100$              & 160 $\pm$ 150       & 200 $\pm$ 200       &                    &             \\                                                                                                            
                                      & \foii          & $<100$              & $<100$              & $<100$              & 150 $\pm$ 150       & 200 $\pm$ 200       &                    &             \\
                                      & \fcliii        & $<100$              &   --                &   --                & 100:                & --                  &                    &             \\
          $T_\mathrm{e}$ (K)          & \foiii$^{\rm a}$        & {\bf 10900 $\pm$ 300}&{\bf 9900 $\pm$ 600} &{\bf 10900 $\pm$ 950$^{\rm c}$}&{\bf 10300 $\pm$ 300}&{\bf 11800 $\pm$ 700}&                    &             \\
                                      & \fnii$^{\rm b}$          & {\bf 10900 $\pm$ 700}&{\bf 9600 $\pm$ 1000}&{\bf 10800 $\pm$ 950}&{\bf 11400 $\pm$ 700}&{\bf 11800 $\pm$ 1900}&                   &              \\
                                      & \foii          & 12300 $\pm$ 700     & 12100 $\pm$ 500     & 15700 $\pm$ 900     &  21300 $\pm$ 2000   &  13100 $\pm$ 700    &                    &             \\   
                                      & \fsii          & 12100 $\pm$ 1500    &        --           & 7300 $\pm$ 1100     &       --            &         --          &                    &             \\
         \hline
         \multicolumn{9}{l}{$^{\rm a}$ $T$(high) = $T_\mathrm{e}$(\foiii) in all M\,33 {\hii} regions.}\\ 
         \multicolumn{9}{l}{$^{\rm b}$ $T$(low) = $T_\mathrm{e}$(\fnii) in all M\,33 {\hii} regions.}\\ 
         \multicolumn{9}{l}{$^{\rm c}$ $T_\mathrm{e}$(\foiii) determined from $T_\mathrm{e}$(\fnii) and assuming equation no. 2.}\\

 \end{tabular}
 \end{scriptsize}
\end{table*}

For the study of \hii\ regions in NGC\,300, we assumed a two-zone electron temperature scheme, where the high-ionization zone is characterized by a $T$(high) that is the weighted mean of $T_\mathrm{e}$(\foiii) and $T_\mathrm{e}$(\fsiii) whereas the weighted mean of $T_\mathrm{e}$(\fnii) and $T_\mathrm{e}$(\foii) represents the $T$(low) for the low-ionization zone. For the most external \hii\ regions in NGC\,300, the auroral {\fnii} 5755 \AA\ line was not detected or it was extremely faint and its flux was not reliable. In such case, we only used the {\foii} diagnostic to compute $T$(low).  We did not include $T_\mathrm{e}$({\fsii}) in the average because it usually gives electron temperatures lower than the other indicators. This difference has been previously reported in the literature and may be due to different reasons \citep[e.g.][]{2009ApJ...700..654E}. For M\,33, we also adopted a two-zone scheme 
but, in this case, we did not determine $T_\mathrm{e}$(\fsiii) due to the shorter wavelength coverage of the OSIRIS observations, therefore we assumed $T_\mathrm{e}$(\foiii) as representative of $T$(high). In the \hii\ regions B2011b15 and LGC\,\hii3, the \foiii\ 4363 \AA\ line was not detected and we derived $T$(high) through the relation obtained by \citet{2009ApJ...700..654E}:

\begin{equation} 
T_\mathrm{e}(\mathrm{\foiii})  = (T_\mathrm{e}(\mathrm{\fnii}) - 3050\ \mathrm{K}) / 0.71
\end{equation}
 
For the objects of M\,33, we adopted $T_\mathrm{e}$(\fnii) as representative of $T$(low).  $T_\mathrm{e}$(\foii) was not used because the trans-auroral \foii\ 7320+30 \AA\ lines are very much affected by sky emission features due to the lower spectral resolution of the M\,33 spectra. We do not use $T_\mathrm{e}$({\fsii}) for the same reason argued for NGC\,300.

The uncertainties on the temperatures were computed through Monte Carlo simulations following a method similar to that explained in \citet{2014ApJ...784..173D}. We generated 500 random values for each line intensity assuming a Gaussian distribution with a sigma equal to the associated uncertainty. We checked that for a higher number of Monte Carlo simulations the errors in the computed quantities remain constant. We calculated new simulated values of $T_\mathrm{e}$ and $n_\mathrm{e}$ (and the ionic and total abundances calculated in Section~\ref{sec:abundances}). The estimated errors correspond to a deviation of 68\% (equivalent to one standard deviation) centered in the mode of the distribution. We decided to discard the standard deviation as representative of the error because the new distributions lost the symmetry property and in this case, the standard deviation is not a good indicator.

\section{Chemical abundances}\label{sec:abundances}

The deep spectra of our eighteen {\hii} regions permit to determine abundances for several ions from CELs and, in some cases, also from RLs. These abundance calculations are done following the direct method, i. e. they are based on determinations of $T_{\rm e}$ from line intensity ratios obtained from the spectra. Total abundances are also derived for some elements.

\subsection{Ionic abundances from CELs}\label{sec:ionicCELs}

We derived ionic abundances of N$^+$, O$^+$, O$^{2+}$, S$^+$, S$^{2+}$, Ar$^{2+}$ and Fe$^{2+}$ from CELs for all observed {\hii} regions, Ne$^{2+}$ for all objects except B2011b5 in M\,33, and Cl$^{2+}$, and Ar$^{3+}$ for only some \hii\ regions. All the computations were made with PyNeb. To make a proper comparison with the results obtained by \citet{2009ApJ...700..309B} in NGC\,300, we used the same atomic dataset than those authors (see Table~\ref{tab:atomicData}). For consistency, this atomic dataset has been also used for M\,33. We assumed a two zone scheme adopting  $T$(low) for ions with low ionization potential (N$^+$, O$^+$, S$^+$, and Fe$^{2+}$) and $T$(high) for the high ionization potential ions (O$^{2+}$, Ne$^{2+}$, S$^{2+}$, Cl$^{2+}$, Ar$^{2+}$ and Ar$^{3+}$). As it was said before, we assumed a canonical value of  $n_\mathrm{e}$ = 100 cm$^{-3}$ for all the objects with $n_\mathrm{e}$ below 100 cm$^{-3}$. 
Tables~\ref{tab:ion_abun} and \ref{tab:ion_abun_m33} show the ionic abundances and their uncertainties that have been computed through Monte Carlo simulations taking into account the uncertainties in line fluxes, $n_\mathrm{e}$, and $T_\mathrm{e}$ (see Section \ref{sec:physicalcondition}). We have considered two sets of abundances, one for $t^2 = 0$ and other for $t^2 > 0$ (see Section~\ref{t2}).

\subsection{Ionic abundances from RLs}\label{sec:ionicRLs}

We detect several {\hei} emission lines in the spectra of our {\hii} regions. Although these lines arise mainly from recombination, those belonging to triplet transitions can be affected by collisional excitation and self-absorption effects. In order to minimize these effects, we use the recent effective recombination coefficient computations by \citet{2012MNRAS.425L..28P, 2013MNRAS.433L..89P} for {\hei} lines where both collisional contribution effects and the optical depth in the triplet lines are included. The final adopted He$^{+}$/H$^{+}$ ratio is the weighted average of the brightest {\hei} emission lines: 4471, 5876, and 6678 \AA\ (1:3:1; according to the intrinsic intensity ratios of the three lines). The He$^+$/H$^+$ ratios are shown in Tables~\ref{tab:ion_abun} and \ref{tab:ion_abun_m33}.
\begin{figure*}
 \centering
  \includegraphics[width=0.49\textwidth]{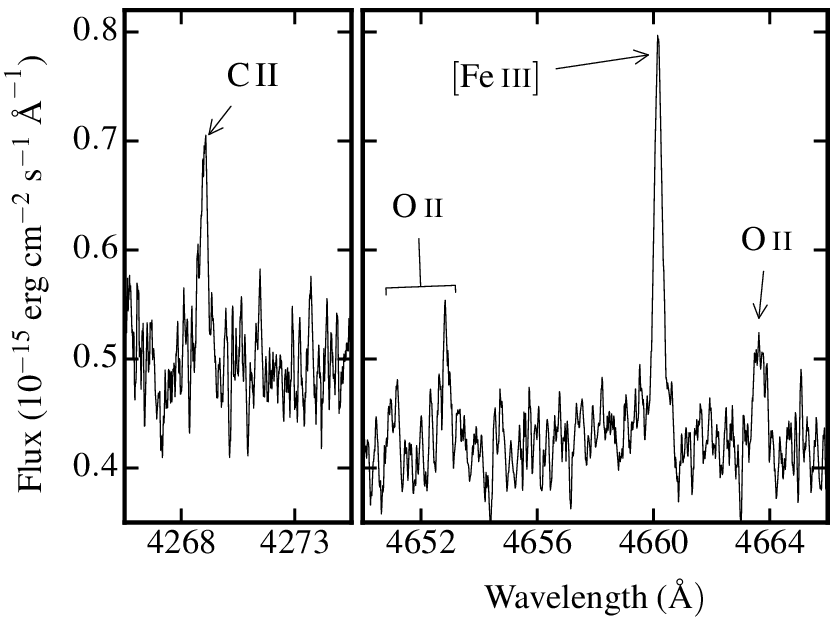}
  \includegraphics[width=0.49\textwidth]{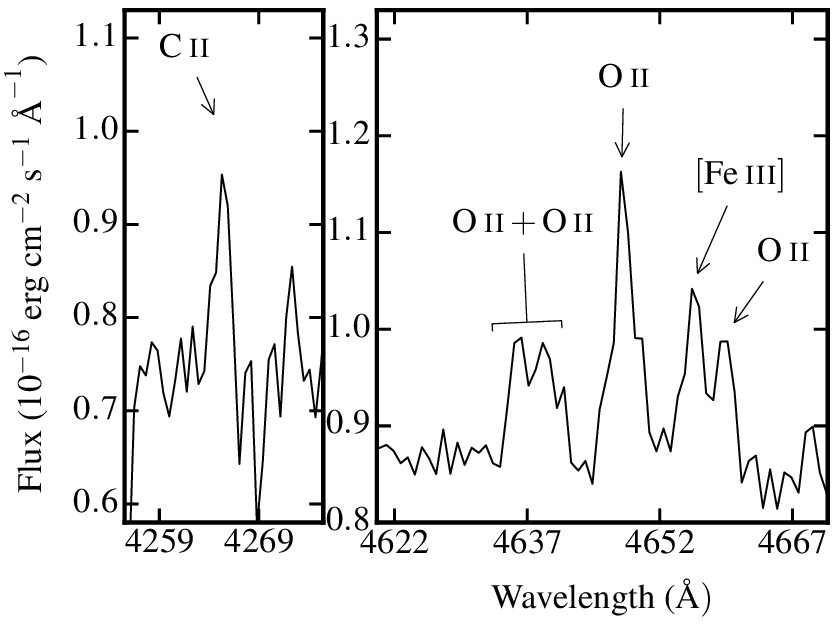}
  \caption{Sections of the UVES spectra of the {\hii} region R14 of NGC\,300 (left) and of the GTC spectra of the {\hii} region NGC\,588 of M\,33 (right), which show the recombination line of {\cii} 4267 \AA\ (left panels) and the recombination lines of multiplet 1 of {\oii} about 4650 \AA\ (right panels).} 
  \label{fig:spec_CyO}
\end{figure*}

The main aim of this work is the study of the C and O radial gradients in NGC\,300 and M\,33 using RLs. The \cii\ $\lambda$4267 was detected in four of the seven \hii\ regions observed in NGC\,300 and in four of the eleven \hii\ regions in M\,33. In addition, \oii\ multiplet 1 around 4650 \AA\ RLs were detected in the same four \hii\ regions in NGC\,300 but only in two in M\,33.

In Figure \ref{fig:hiiregions} we mark with double circles the \hii\ regions where we detect \cii\ RL. These \hii\ regions correspond to R2, R14, R20, and R23 as identified by \citet{2009ApJ...700..309B} in NGC\,300 and with B2011\,b5, BCLMP\,29, NGC\,588 and IC\,132 in M\,33. Although the uncertainties of the detected  \cii\ and  \oii\ lines are large owing to their intrinsic faintness, we could derive the ionic abundance ratios $\mathrm{C^{2+}}$/$\mathrm{H^{+}}$ in all these {\hii} regions and $\mathrm{O^{2+}}$/$\mathrm{H^{+}}$ in six of them. Figure~\ref{fig:spec_CyO} shows two sections of the spectrum of the \hii\ region R14 of NGC\,300 and the \hii\ region NGC\,588 of M\,33, centered in \cii\ 4267 \AA\ RL (left panels) and in multiplet 1 of \oii\ around 4650 \AA\ lines (right panels).

Let $I$($\lambda$) be the intensity of an RL of an element X, which is $i$ times ionized; the abundance of the ionization state $i + 1$  of element X is given by the following expression: 

\begin{equation} 
\frac{N(\mathrm{X}^{i+1})}{N(\mathrm{H}^+)} = \frac{\lambda({\mathrm\AA})}{4861} \frac{\alpha_{\mathrm{eff}}(\mathrm{H\beta}) }{\alpha_{\mathrm{eff}}(\lambda)} \frac{I(\lambda)}{I(\mathrm{H\beta})}\label{eq:abun_RLs}
\end{equation}
where $\lambda$(\AA) is the wavelength of the RL of the ion $\mathrm{X}^{i+}$,  $\alpha_{\mathrm{eff}}(\mathrm{H\beta})$ and $\alpha_{\mathrm{eff}}(\lambda)$ are the effective recombination coefficients for $\mathrm{H\beta}$ and the RL, respectively, and $I(\lambda)$/$I(\mathrm{H\beta})$ is the intensity line ratio of the RL with respect to H$\beta$.

We computed the $\mathrm{C^{2+}}$ abundance from the measured flux of the \cii\ 4267 \AA\ RL, $T_\mathrm{e}$(high), and the \cii\ effective recombination coefficients calculated by \citet{2000A&AS..142...85D} for case B. 

We also computed $\mathrm{O^{2+}}$ abundances from RLs belonging to multiplet 1 of \oii. As \citet{2003ApJ...595..247R} pointed out, the relative populations of the levels do not follow local thermodinamic equilibrium (LTE)  for densities $n_\mathrm{e}<10^4\ \mathrm{cm}^{-3}$. We used the prescriptions of \citet{2005RMxAC..23....9P} to calculate the appropriate corrections for the relative strengths between the individual \oii\ lines and then use $T_\mathrm{e}$(high), and the \oii\ effective recombination coefficients for case B and assuming LS coupling calculated by \citet{1994A&A...282..999S} . The final  $\mathrm{C^{2+}}$ and $\mathrm{O^{2+}}$ abundances determined from RLs are shown in Table~\ref{tab:abun_RLs}.

\subsection{Total abundances}\label{sec:total}

To compute the elemental or total abundances we have to correct for the unseen ionization stages of each element. Then, we need to adopt a set of ionization correction factors (ICFs). 

Since we do not detect {\heii} lines in any of the spectra of our \hii\ regions, we consider that there is no O$^{3+}$ in these objects. Therefore, no ICF is needed to compute the total O abundance from CELs. The O/H ratios are calculated simply adding O$^+$ and O$^{2+}$ abundances.

For He, we have considered the ICF given by \citet{1992RMxAA..24..155P}. To compute the total abundances of N, Ne, S, Cl and Ar from CELs, we adopted the ICFs proposed by \citet{2006A&A...448..955I} for extragalactic {\hii} regions and based on 
photoionization models. These authors propose different ICFs depending on the element and the metallicity of the \hii\ region. We used the expressions for high metallicity (i.e. $\mathrm{Z_{\odot}} > 0.2$) in all our regions. We take into account the presence of {\fariv} lines in the spectrum of R2, NGC\,588 and IC132 using the ICF recommended by \citet{2006A&A...448..955I} for this case. For Fe, we adopted the ICF scheme suggested by \citet{2005ApJ...626..900R} which is based on the computed Fe$^{2+}$/H$^+$ ratio and two different ICFs: one based on photoionization models (their equation 2) and another one based on an observational fit (their equation 3). 

In the case of C, we used the ICF derived by \citet{1999ApJ...513..168G} from photoionization models to correct from the unseen ionization stages of this element. In addition, we also derived the total O abundance using the O$^{2+}$/H$^+$ ratio derived from 
RLs and assuming the O$^+$/O$^{2+}$ ratio obtained from CELs to correct for the contribution of the O$^+$/H$^+$ ratio.  

Tables \ref{tab:ion_abun} and \ref{tab:ion_abun_m33} present the total abundances obtained from heavy element CELs -- assuming $t^2 = 0$ and $t^2 > 0$ (see section~\ref{t2}) -- and RLs of {\hei}. Table \ref{tab:abun_RLs} shows the total abundances of C and O determined from RLs .
\begin{landscape}
   \begin{table}
   \begin{minipage}{230mm}
       \caption{Ionic and total abundances in units of $12+\log(\mathrm{X^{n+}/H^+})$ of \hii\ regions in NGC\,300.$^{\rm a}$}
       \label{tab:ion_abun}
       \begin{tabular}{lccccccccccc}
           \hline
                              & R76a           & \multicolumn{2}{c}{R20}           & \multicolumn{2}{c}{R23}           & \multicolumn{2}{c}{R14}           & R5              & R27             & \multicolumn{2}{c}{R2}             \\
                              &                &                 & $t^2$ = $0.035 $&                 & $t^2$ = $0.035 $&                 & $t^2$ = $0.045$ &                 &                 &                 & $t^2$ = $0.089 $ \\
                              & $t^2$ = 0      &  $t^2$ = 0      &  $ \pm 0.021$   &  $t^2$ = 0      &    $\pm 0.019$  &  $t^2$ = 0      &  $ \pm 0.019$   &  $t^2$ = 0      & $t^2$ = 0       &  $t^2$ = 0      &   $\pm 0.028$    \\
            \hline   
            \\                          
            \multicolumn{12}{c}{Ionic abundances} \\
            \\
           $\mathrm{He^{+}}$   & $10.78 \pm 0.05$& \multicolumn{2}{c}{$10.96 \pm 0.03$}  & \multicolumn{2}{c}{$ 10.92 \pm 0.02 $} & \multicolumn{2}{c}{$ 10.89 \pm 0.03 $} & $ 10.93 \pm 0.03$ & $10.72 \pm 0.07$ & \multicolumn{2}{c}{$10.93 \pm 0.03$} \\
           $\mathrm{N^{+}}$   & $7.48 \pm 0.07$& $6.93 \pm 0.05$ & $7.08 \pm 0.06$ & $6.82 \pm 0.05$ & $6.95 \pm 0.06$ & $6.86 \pm 0.04$ & $7.05 \pm 0.05$ & $6.90 \pm 0.04$ & $6.85 \pm 0.06$ & $6.19 \pm 0.05$ & $6.44 \pm 0.07$\\
           $\mathrm{O^{+}}$   & $8.40 \pm 0.13$& $8.02 \pm 0.08$ & $8.14 \pm 0.17$ & $7.85 \pm 0.09$ & $7.96 \pm 0.15$ & $7.89 \pm 0.07$ & $8.04 \pm 0.15$ & $8.17 \pm 0.07$ & $8.02 \pm 0.08$ & $7.70 \pm 0.06$ & $7.90 \pm 0.12$\\
           $\mathrm{O^{2+}}$  & $7.20 \pm 0.43$& $8.23 \pm 0.08$ & $8.52 \pm 0.09$ & $8.23 \pm 0.08$ & $8.52 \pm 0.09$ & $8.07 \pm 0.07$ & $8.43 \pm 0.07$ & $7.87 \pm 0.11$ & $7.52 \pm 0.28$ & $8.03 \pm 0.07$ & $8.43 \pm 0.07$\\
           $\mathrm{Ne^{2+}}$ & $5.95 \pm 0.26$& $7.42 \pm 0.09$ & $7.73 \pm 0.12$ & $7.49 \pm 0.09$ & $7.81 \pm 0.12$ & $7.25 \pm 0.07$ & $7.64 \pm 0.10$ & $7.07 \pm 0.13$ & $6.73 \pm 0.37$ & $7.36 \pm 0.07$ & $7.79 \pm 0.09$\\
           $\mathrm{S^{+}}$   & $6.44 \pm 0.07$& $5.88 \pm 0.05$ & $6.03 \pm 0.06$ & $5.92 \pm 0.05$ & $6.05 \pm 0.06$ & $5.95 \pm 0.04$ & $6.14 \pm 0.05$ & $6.19 \pm 0.04$ & $6.41 \pm 0.05$ & $5.63 \pm 0.04$ & $5.88 \pm 0.06$\\
           $\mathrm{S^{2+}}$  & $6.81 \pm 0.59$& $6.87 \pm 0.06$ & $7.13 \pm 0.08$ & $6.83 \pm 0.05$ & $7.10 \pm 0.08$ & $6.80 \pm 0.05$ & $7.12 \pm 0.07$ & $6.58 \pm 0.09$ & $6.64 \pm 0.16$ & $6.33 \pm 0.06$ & $6.70 \pm 0.08$\\
           $\mathrm{Cl^{2+}}$ & --             & $4.85 \pm 0.18$ & $5.12 \pm 0.38$ & $4.86 \pm 0.14$ & $5.13 \pm 0.32$ & $4.89 \pm 0.14$ & $5.23 \pm 0.22$ & $4.51 \pm 0.40$ &--               & $4.66 \pm 0.16$ & $5.04 \pm 0.19$\\
           $\mathrm{Ar^{2+}}$ & $5.85 \pm 0.69$& $6.20 \pm 0.06$ & $6.44 \pm 0.07$ & $6.17 \pm 0.06$ & $6.42 \pm 0.06$ & $6.14 \pm 0.05$ & $6.44 \pm 0.06$ & $5.94 \pm 0.08$ & $5.68 \pm 0.22$ & $5.76 \pm 0.05$ & $6.10 \pm 0.07$\\
           $\mathrm{Ar^{3+}}$ & --             & --              & --              & --              & --              & --              & --              & --              & --              & $4.47 \pm 0.26$ & $4.88 \pm 0.35$\\
           $\mathrm{Fe^{2+}}$ & $5.56 \pm 0.37$& $5.10 \pm 0.20$ & $5.39 \pm 0.25$ & $5.28 \pm 0.14$ & $5.59 \pm 0.16$ & $4.98 \pm 0.13$ & $5.34 \pm 0.24$ & $5.50 \pm 0.28$ & $6.11 \pm 0.36$ & $4.63 \pm 0.37$ & $5.05 \pm 0.78$\\
           \\
           \multicolumn{12}{c}{Total abundances}\\      
           \\                                                                                                   
           $\mathrm{He}$   & $10.99 \pm 0.06$& \multicolumn{2}{c}{$11.00 \pm 0.03$} & \multicolumn{2}{c}{$ 10.98 \pm 0.03$} & \multicolumn{2}{c}{$10.95 \pm 0.03$} & $ 11.11 \pm 0.05 $ & $10.96 \pm 0.11$ & \multicolumn{2}{c}{$ 11.01 \pm 0.03 $} \\
           $\mathrm{N}$ & $ 7.52 \pm 0.09$    & $7.40 \pm 0.05$ & $7.64 \pm 0.02$ & $7.38 \pm 0.05$ & $7.62 \pm 0.02$ & $7.33 \pm 0.04$ & $7.62 \pm 0.02$  &$7.16 \pm 0.05$ & $7.03 \pm 0.11$ & $6.72 \pm 0.05$ & $7.09 \pm 0.04$ \\
           $\mathrm{O}$ & $8.42 \pm 0.13$     & $8.44 \pm 0.06$ & $8.67 \pm 0.12$ & $8.38 \pm 0.06$ & $8.63 \pm 0.10$ & $8.29 \pm 0.05$ & $8.57 \pm 0.09$  &$8.35 \pm 0.06$ & $8.14 \pm 0.10$ & $8.19 \pm 0.05$ & $8.54 \pm 0.08$ \\
           $\mathrm{Ne}$ & $6.95 \pm 0.28$    & $7.59 \pm 0.07$ & $7.84 \pm 0.14$ & $7.60 \pm 0.07$ & $7.88 \pm 0.14$ & $7.42 \pm 0.06$ & $7.74 \pm 0.12$  &$7.45 \pm 0.09$ & $7.22 \pm 0.20$ & $7.48 \pm 0.06$ & $7.86 \pm 0.10$ \\ 	
           $\mathrm{S}$ & $6.94 \pm 0.14$     & $6.95 \pm 0.07$ & $7.24 \pm 0.06$ & $6.95 \pm 0.07$ & $7.27 \pm 0.05$ & $6.88 \pm 0.05$ & $7.24 \pm 0.05$  &$6.71 \pm 0.07$ & $6.82 \pm 0.11$ & $6.47 \pm 0.06$ & $6.89 \pm 0.06$ \\
           $\mathrm{Cl}$ & --                 & $4.91 \pm 0.18$ & $5.18 \pm 0.42$ & $4.92 \pm 0.15$ & $5.22 \pm 0.37$ & $4.96 \pm 0.14$ & $5.29 \pm 0.22$  &$4.64 \pm 0.40$ & --              & $4.71 \pm 0.16$ & $5.13 \pm 0.18$ \\
           $\mathrm{Ar}$ & $5.96 \pm 0.29$    & $6.23 \pm 0.05$ & $6.46 \pm 0.07$ & $6.19 \pm 0.06$ & $6.45 \pm 0.06$ & $6.17 \pm 0.05$ & $6.46 \pm 0.06$  &$6.01 \pm 0.07$ & $5.76 \pm 0.20$ & $5.79 \pm 0.05$  & $6.13 \pm 0.08$\\
           $\mathrm{Fe^b}$ & $-^{\mathrm{c}}$ & $5.44 \pm 0.20$ & $5.74 \pm 0.25$ & $5.63 \pm 0.14$ & $5.97 \pm 0.17$ & $5.32 \pm 0.13$ & $5.70 \pm 0.24$  &$-^{\mathrm{c}}$& $-^{\mathrm{c}}$& $4.98 \pm 0.40$  & $5.43 \pm 0.59$\\
           $\mathrm{Fe^d}$ & $5.64 \pm 0.41$  & $5.45 \pm 0.21$ & $5.84 \pm 0.21$ & $5.73 \pm 0.14$ & $6.17 \pm 0.12$ & $5.32 \pm 0.13$ & $5.80 \pm 0.20$  &$5.65 \pm 0.28$ & $6.22 \pm 0.38$ & $5.05 \pm 0.38$  & $5.61 \pm 0.61$\\
           
           \hline\\
           \multicolumn{12}{l}{$^{\rm a}$ C$^{++}$ and O$^{++}$ abundances  determined from RLs are given in Table~\ref{tab:abun_RLs}.}\\
           \multicolumn{12}{l}{$^{\rm b}$ Calculated using equation 3 of \citet{2005ApJ...626..900R}: least-squares linear fit to observations.}\\
           \multicolumn{12}{l}{$^{\rm c}$ Equation 3 of \citet{2005ApJ...626..900R} not applicable.}  \\
           \multicolumn{12}{l}{$^{\rm d}$ Calculated using equation 2 of \citet{2005ApJ...626..900R}: least-squares fit to photoionization models results.}\\
     \end{tabular}
  \end{minipage}
  \end{table}
\end{landscape}

\begin{landscape}
   \begin{table}
   \begin{scriptsize}
   \begin{minipage}{230mm}
       \caption{Ionic and total abundances in units of $12+\log(\mathrm{X^{n+}/H^+})$ of \hii\ regions in M\,33.$^{\rm a}$}
       \label{tab:ion_abun_m33}
       \begin{tabular}{l c c c c c c c c c c c c c}
           \hline
                                  &B2011b5         & BCLMP\,29       &B2011b15         & BCLMP\,88w      & IC\,131         & BCLMP\,290      & \multicolumn{2}{c}{NGC\,588}        & BCLMP\,626      & LGC\,\hii3         & \multicolumn{2}{c}{IC\,132}   &  LGC\,\hii11    \\
                                  &                &                 &                 &                 &                 &                 &                 &$t^2$ = $0.059$    &                 &                 &                 &$t^2$ = $0.038$ &                 \\
                                  &$t^2$ = 0       &  $t^2$ = 0      &$t^2$ = 0        &$t^2$ = 0        &$t^2$ = 0        & $t^2$ = 0       &$t^2$ = 0        &$\pm 0.018$        &$t^2$ = 0        &$t^2$ = 0        &$t^2$ = 0        &$\pm 0.019$     & $t^2$ = 0       \\
            \hline                                                                                                                                                                                                                                                                                                                                                                                                      
            \\
            \multicolumn{14}{c}{Ionic abundances} \\                                                                                                                                                                                                                                                                                                                                                                                   
           \\
           $\mathrm{He^{+}}$      & $10.99 \pm 0.05$ & $10.88 \pm 0.03$ & $10.75 \pm 0.05$ & $10.95 \pm 0.06$ & $10.93 \pm 0.03$ & $10.88 \pm 0.03$ & \multicolumn{2}{c}{$10.88 \pm 0.03$} & $10.87 \pm 0.05$ & $10.93 \pm 0.03$ & \multicolumn{2}{c}{$10.94 \pm 0.03$} & $10.92 \pm 0.04$   \\
           $\mathrm{N^{+}}$      & $7.26 \pm 0.14$ & $7.42 \pm 0.06$ & $7.26 \pm 0.18$ & $6.93 \pm 0.24$ & $6.81 \pm 0.23$ & $6.83 \pm 0.09$ & $6.12 \pm 0.11$ & $6.27 \pm 0.14$ & $6.78 \pm 0.19$ & $6.55 \pm 0.10$ & $6.03 \pm 0.11$ & $6.12 \pm 0.12$& $6.06 \pm 0.16$   \\ 
           $\mathrm{O^{+}}$      & $8.03 \pm 0.18$ & $8.33 \pm 0.13$ & $8.25 \pm 0.29$ & $7.95 \pm 0.39$ & $7.98 \pm 0.32$ & $8.00 \pm 0.15$ & $7.38 \pm 0.15$ & $7.55 \pm 0.20$ & $8.02 \pm 0.30$ & $7.66 \pm 0.14$ & $7.10 \pm 0.16$ & $7.20 \pm 0.20$& $7.36 \pm 0.22$   \\ 
           $\mathrm{O^{2+}}$     & $8.28 \pm 0.15$ & $7.72 \pm 0.16$ & $7.94 \pm 0.39$ & $8.19 \pm 0.18$ & $8.25 \pm 0.09$ & $8.11 \pm 0.08$ & $8.14 \pm 0.07$ & $8.42 \pm 0.08$ & $7.78 \pm 0.12$ & $7.89 \pm 0.16$ & $8.25 \pm 0.09$ & $8.44 \pm 0.09$& $7.85 \pm 0.12$   \\ 
           $\mathrm{Ne^{2+}}$    &    --           & $6.82 \pm 0.20$ & $7.36 \pm 0.49$ & $7.46 \pm 0.22$ & $7.58 \pm 0.10$ & $7.31 \pm 0.09$ & $7.41 \pm 0.07$ & $7.70 \pm 0.08$ & $6.91 \pm 0.15$ & $6.92 \pm 0.21$ & $7.61 \pm 0.09$ & $7.81 \pm 0.11$& $6.06 \pm 0.15$   \\ 
           $\mathrm{S^{+}}$      & $6.09 \pm 0.10$ & $6.33 \pm 0.06$ & $6.47 \pm 0.17$ & $5.94 \pm 0.24$ & $6.09 \pm 0.20$ & $5.89 \pm 0.08$ & $5.50 \pm 0.10$ & $5.66 \pm 0.11$ & $6.05 \pm 0.17$ & $5.75 \pm 0.08$ & $5.30 \pm 0.09$ & $5.38 \pm 0.12$& $5.43 \pm 0.14$   \\ 
           $\mathrm{S^{2+}}$     & - & $6.60 \pm 0.21$ & $6.86 \pm 0.53$ & $6.74 \pm 0.24$ & $6.71 \pm 0.27$ & $6.79 \pm 0.10$ & $6.44 \pm 0.09$ & $6.73 \pm 0.09$ & $6.51 \pm 0.17$ & $6.41 \pm 0.22$ & $6.52 \pm 0.10$ & $6.72 \pm 0.11$& $6.26 \pm 0.18$   \\ 
           $\mathrm{Cl^{2+}}$    & $5.15 \pm 0.13$ & $4.81 \pm 0.16$ &     --          & $4.90 \pm 0.20$ & $4.80 \pm 0.12$ & $4.92 \pm 0.08$ & $4.60 \pm 0.06$ & $4.86 \pm 0.08$ & $4.69 \pm 0.21$ & $4.55 \pm 0.16$ & $4.66 \pm 0.07$ & $4.84 \pm 0.08$& $4.59 \pm 0.17$   \\ 
           $\mathrm{Ar^{2+}}$    & $6.36 \pm 0.13$ & $6.04 \pm 0.12$ & $6.05 \pm 0.28$ & $6.17 \pm 0.16$ & $6.15 \pm 0.11$ & $6.17 \pm 0.06$ & $5.89 \pm 0.07$ & $6.12 \pm 0.07$ & $5.85 \pm 0.09$ & $5.92 \pm 0.12$ & $5.93 \pm 0.07$ & $6.09 \pm 0.07$& $5.75 \pm 0.12$   \\ 
           $\mathrm{Ar^{3+}}$    &    --           &     --          &      --         &        --       &     --          &      --         & $4.77 \pm 0.09$ & $5.05 \pm 0.10$ &    --           & --              & $4.93 \pm 0.09$ & $5.12 \pm 0.09$&  --               \\ 
           $\mathrm{Fe^{2+}}$    & $5.51 \pm 0.16$ & $5.38 \pm 0.10$ & $5.63 \pm 0.32$ & $5.52 \pm 0.34$ & $5.30 \pm 0.33$ & $5.23 \pm 0.11$ & $4.43 \pm 0.15$ & $4.70 \pm 0.20$ & $5.54 \pm 0.33$ & $4.99 \pm 0.13$ &  --             &      --        & --                \\                                                                                                                                                                   
           \\
           \multicolumn{14}{c}{Total abundances}\\                                                                                                                                                                                                                                                                                                                                                                                             
           \\
           $\mathrm{He}$          & --- &  $11.09 \pm 0.09$ & $10.90 \pm 0.12$ & $11.01 \pm 0.08$  & $11.01 \pm 0.10$ & $10.92 \pm 0.04$ & \multicolumn{2}{c}{$10.90 \pm 0.03$} & $10.96 \pm 0.07$ & $11.01 \pm 0.04$ & \multicolumn{2}{c}{$10.95 \pm 0.03$} & $10.95 \pm 0.05$   \\ 
           $\mathrm{N}$          &$7.76 \pm 0.10$  & $7.57 \pm 0.07$ & $7.51 \pm 0.22$ & $7.43 \pm 0.16$ & $7.31 \pm 0.11$ & $7.25 \pm 0.05$ & $6.91 \pm 0.06$ & $7.14 \pm 0.06$ & $7.06 \pm 0.10$ & $7.03 \pm 0.11$ & $7.11 \pm 0.08$ & $7.28 \pm 0.03$& $6.68 \pm 0.10$   \\ 
           $\mathrm{O}$          &$8.47 \pm 0.11$  & $8.42 \pm 0.08$ & $8.42 \pm 0.33$ & $8.39 \pm 0.20$ & $8.43 \pm 0.13$ & $8.36 \pm 0.09$ & $8.21 \pm 0.06$ & $8.48 \pm 0.09$ & $8.21 \pm 0.20$ & $8.09 \pm 0.16$ & $8.29 \pm 0.08$ & $8.47 \pm 0.09$& $7.97 \pm 0.10$   \\ 
           $\mathrm{Ne}$         &  --             & $7.37 \pm 0.14$ & $7.74 \pm 0.45$ & $7.60 \pm 0.25$ & $7.71 \pm 0.15$ & $7.50 \pm 0.08$ & $7.44 \pm 0.06$ & $7.71 \pm 0.10$ & $7.25 \pm 0.19$ & $7.07 \pm 0.20$ & $7.59 \pm 0.08$ & $7.79 \pm 0.12$& $7.05 \pm 0.13$   \\      
           $\mathrm{S}$          & -  & $6.76 \pm 0.15$ & $6.98 \pm 0.43$ & $6.83 \pm 0.24$ & $6.85 \pm 0.24$ & $6.85 \pm 0.11$ & $6.71 \pm 0.13$ & $7.05 \pm 0.02$ & $6.62 \pm 0.14$ & $6.53 \pm 0.19$ & $7.01 \pm 0.20$ & $7.28 \pm 0.03$& $6.41 \pm 0.18$   \\ 
           $\mathrm{Cl}$         &$5.20 \pm 0.13$  & $4.97 \pm 0.14$ &    --           & $4.96 \pm 0.21$ & $4.86 \pm 0.16$ & $4.99 \pm 0.07$ & $4.75 \pm 0.11$ & $5.07 \pm 0.02$ & $4.81 \pm 0.21$ & $4.61 \pm 0.15$ & $5.05 \pm 0.18$ & $5.30 \pm 0.01$& $4.66 \pm 0.17$   \\ 
           $\mathrm{Ar}$         &$6.38 \pm 0.13$  & $6.13 \pm 0.11$ & $6.11 \pm 0.27$ & $6.20 \pm 0.17$ & $6.17 \pm 0.12$ & $6.20 \pm 0.06$ & $5.92 \pm 0.06$ & $6.15 \pm 0.07$ & $5.91 \pm 0.11$ & $5.95 \pm 0.12$ & $5.98 \pm 0.07$ & $6.14 \pm 0.06$& $5.77 \pm 0.12$   \\ 
           $\mathrm{Fe^b}$       &$5.86 \pm 0.17$  & $-^{\mathrm{c}}$&$-^{\mathrm{c}}$ & $5.86 \pm 0.34$ & $5.64 \pm 0.31$ & $5.57 \pm 0.11$ & $4.86 \pm 0.11$ & $5.17 \pm 0.18$ &$-^{\mathrm{c}}$ & $5.33 \pm 0.13$ &   --            &    --          &     --            \\ 
           $\mathrm{Fe^d}$       &$5.89 \pm 0.14$  & $5.47 \pm 0.10$ & $5.78 \pm 0.34$ & $5.90 \pm 0.19$ & $5.68 \pm 0.22$ & $5.53 \pm 0.07$ & $5.15 \pm 0.09$ & $5.52 \pm 0.13$ & $5.71 \pm 0.28$ & $5.36 \pm 0.13$ &   --            &    --          &     --            \\ 
           
           \hline\\          
           \multicolumn{14}{l}{$^{\rm a}$ C$^{++}$ and O$^{++}$ abundances determined from RLs are given in Table~\ref{tab:abun_RLs}.}\\
           \multicolumn{14}{l}{$^{\rm b}$ Calculated using equation 3 of \citet{2005ApJ...626..900R}: least-squares linear fit to observations.}\\
           \multicolumn{14}{l}{$^{\rm c}$ Equation 3 of \citet{2005ApJ...626..900R} not applicable.}  \\
           \multicolumn{14}{l}{$^{\rm d}$ Calculated using equation 2 of \citet{2005ApJ...626..900R}: least-squares fit to photoionization models results.}\\
     \end{tabular}
  \end{minipage}
  \end{scriptsize}
  \end{table}
\end{landscape}

\begin{table*}
  \begin{minipage}{140mm}
  \caption{Abundances from RLs in units of $12+\log(\mathrm{X^{n+}/H^+})$.}
  \label{tab:abun_RLs}
    \begin{tabular}{l c c c c}
       \hline
                         & \multicolumn{4}{c}{NGC\,300}\\                                                                                                 
       & R20 & R23 & R14 & R2 \\
       \hline
       \multicolumn{5}{c}{Ionic abundances} \\
       \\
       $\mathrm{C^{++}}$ &$8.30 \pm 0.19$ &$8.13 \pm 0.19$   &  $8.08 \pm 0.18$ & $7.90 \pm 0.19$ \\
       $\mathrm{O^{++}}$ &$8.48 \pm 0.28$ &$8.54 \pm 0.22$   &  $8.40 \pm 0.25$   & $8.40 \pm 0.16$  \\
       $\mathrm{ADF(O^{++})}$ & 0.25 $\pm$ 0.18 & 0.31 $\pm$ 0.24 & 0.33 $\pm$ 0.19 & 0.37 $\pm$ 0.21 \\
       \\
       \multicolumn{5}{c}{Total abundances}\\
       \\
       $\mathrm{C}$ &$8.47 \pm 0.19$ &$8.26 \pm 0.19$&  $8.25 \pm 0.18$& $8.04 \pm 0.19$ \\ 
       $\mathrm{O}$ &$8.69 \pm 0.11$ &$8.70 \pm 0.13$&  $8.62 \pm 0.11$& $8.57 \pm 0.12$ \\
       C/O & $-$0.22 $\pm$ 0.22 & $-$0.43 $\pm$ 0.23 & $-$0.36 $\pm$ 0.21 & $-$0.52 $\pm$ 0.22 \\
       \hline 
                              &\multicolumn{4}{c}{M\,33} \\                                                                                                      
                              & B2011b5            & BCLMP29           & NGC\,588             & IC132                \\
       \hline                                                                                                        
       \multicolumn{5}{c}{Ionic abundances} \\                                                                       
       \\                                                                                                            
       $\mathrm{C^{++}}$      &$8.32 \pm 0.19$     &$8.19 \pm 0.31$    & $7.95 \pm 0.17$      &  $8.11 \pm 0.12$     \\
       $\mathrm{O^{++}}$      &     --             &        --         & $8.40 \pm 0.10$      &  $8.45 \pm 0.12$     \\
       $\mathrm{ADF(O^{++})}$ &     --             &        --         & 0.26 $\pm$ 0.17        &  0.20 $\pm$ 0.15        \\
       \\                                                                                                            
       \multicolumn{5}{c}{Total abundances}\\                                                                        
       \\                                                                                                            
       $\mathrm{C}$           &$8.48 \pm 0.19$     &$8.73 \pm 0.32$    & $8.02 \pm 0.17$      &  $8.14 \pm 0.12$     \\ 
       $\mathrm{O}$           &     --             &     --            & $8.49 \pm 0.09$      &  $8.47 \pm 0.12$     \\
       C/O                    &     --             &     --            & $-$0.47 $\pm$ 0.19   & $-$0.33 $\pm$ 0.15   \\
       \hline 
   \end{tabular}
 \end{minipage}
\end{table*} 

\subsection{Abundance discrepancy and temperature fluctuations}\label{t2}

The standard method for deriving ionic abundances is based on the intensity of CELs. These lines are much brighter than RLs, and therefore easier to detect. However, abundances relative to hydrogen derived from CELs are strongly affected by uncertainties in the determination of $T_\mathrm{e}$. On the other hand, abundances derived from RLs are almost independent on the physical conditions, but in extragalactic \hii\ regions these lines are difficult to observe \citep{2002ApJ...581..241E, 2009ApJ...700..654E, 2014MNRAS.443..624E}. A well-known result is that O$^{2+}$ abundances derived from RLs are systematically a factor between 1.3 and 3  higher than those computed from CELs \citep[e.g.][]{2007ApJ...670..457G}. This dichotomy, the so-called abundance discrepancy problem, is an unsolved issue in nebular astrophysics and several mechanisms have been proposed to explain it. The first hypothesis 
is based on the presence of temperature fluctuations \citep{1969BOTT....5....3P} in the ionized gas of the nebulae. They would affect the abundances derived from CELs but remain unaltered the ones derived from RLs. 
A second explanation assumes the presence of a cold, dense, H-poor (high metallicity) component from which most of the RL emission comes \citep{2005MNRAS.364..687T, 2007A&A...471..193S} and that will affect slightly the abundances derived from CELs and invalidate those obtained from RLs; the responsible of most of the RL emission would be cold metal-rich droplets (the cold component) from supernova ejecta still not mixed with the ambient gas of the \hii\ region (the hot component) where most of the CEL emission would be produced. This is an interesting hypothesis but seems unlikely taking into account that abundance discrepancies are found to be very similar in different galaxies regardless of their masses and star formation rates. Moreover, if this last hypothesis is true, then neither RLs (strongly overestimated abundances), nor CELs (slightly underestimated abundances) would give the ``real'' abundances of the nebula. A last and more recent hypothesis assumes that the energy distribution of electrons in \hii\ regions departs from a Maxwell-Boltzmann one in form of a ``kappa-distribution'' \citep{2012ApJ...752..148N}. This last scenario could be considered as a special case of the temperature fluctuations hypothesis and, also in this case, abundances obtained from RLs would be more representative of the ``real" ones.

We computed $\mathrm{O^{2+}}$/$\mathrm{H^{+}}$ ratios using CELs and RLs in six of the {\hii} regions of our sample, as expected, the abundances obtained from RLs are always higher than those  determined from CELs. We can quantify this difference defining the  abundance discrepancy factor (hereafter ADF) as the difference between the logarithmic abundances derived from RLs and CELs:

\begin{equation}
\mathrm{ADF(X^{i+}) = log(X^{i+}/H^+)_{RLs} - log(X^{i+}/H^+)_{CELs};}
\end{equation}
where $\mathrm{X}^{i+}$ corresponds to the ionization state i of element X. The values of the ADF($\mathrm{O^{2+}}$) found in our \hii\ regions are shown in Table~\ref{tab:abun_RLs} and the weighted mean value is 0.32 $\pm$ 0.06 dex for NGC\,300 and 0.23 $\pm$ 0.04 dex for M\,33. The value obtained in NGC\,300 is close to the upper limit of the typical ADFs reported in the literature for Galactic and extragalactic \hii\ regions \citep[see e.~g.][]{2007ApJ...670..457G, 2009ApJ...700..654E}, which are usually between 0.1 and 0.3 dex, but similar to the values found by \citet{2014MNRAS.443..624E} for a sample of star-forming dwarf galaxies.

If we assume the presence of spatial temperature fluctuations as the cause of the abundance discrepancy, then we have to correct abundances obtained from CELs. Under the temperature fluctuations paradigm, the structure of a gaseous nebula is characterized by two parameters: the average temperature, $T_0$, and the mean square temperature fluctuation, $t^2$ \citep{1967ApJ...150..825P}. For each \hii\ region where we measure \oii\ RLs, we can compute the $T_0$ that generates the same $\mathrm{O^{2+}}$/H$^+$ ratio from CELs and from RLs. Then, with the corresponding value of $t^2$, we can correct the abundances derived from CELs following the formalism firstly developed by \citet{1969BOTT....5....3P}. The new computed ionic and total abundances for $t^2>0$ are presented in Tables~\ref{tab:ion_abun} and \ref{tab:ion_abun_m33}.

\begin{figure*}
 \centering
  \includegraphics[width=0.49\textwidth]{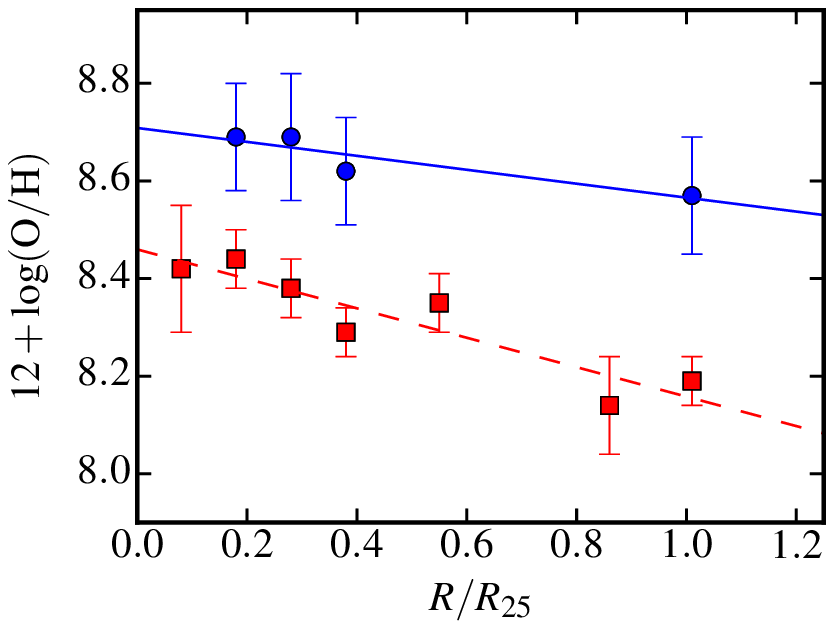}
  \includegraphics[width=0.49\textwidth]{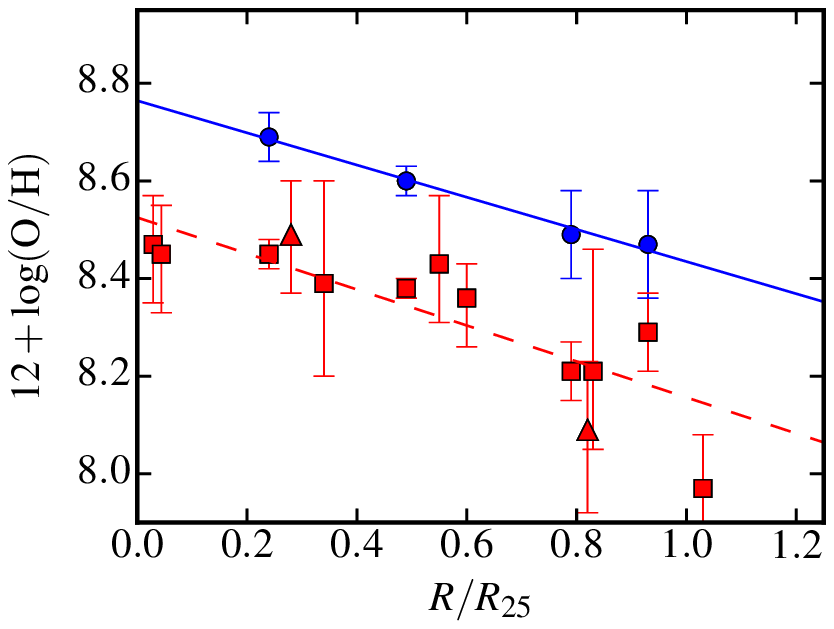}
 \caption{Comparison between the O/H ratios and radial gradients as a function of the fractional galactocentric distance ($R$/$R_{25}$) we obtain for NGC\,300 (left panel) and M\,33 (right panel) from $\mathrm{O^{2+}}$/H$^+$ ratios determined from CELs (red squares) or from RLs (blue circles). The red triangles in M\,33 (right panel) represent the \hii\ regions for which we have estimated $T_\mathrm{e}$(\foiii) from $T_\mathrm{e}$(\fnii) and assuming equation no. 2. We have included the O/H ratios of the two \hii\ regions observed by \citet{2009ApJ...700..654E} in M\,33.}
 \label{fig:DA}
\end{figure*}

\begin{table*}
  \begin{minipage}{140mm}
   \caption{Radial abundance gradients for NGC\,300 and M\,33.}
   \label{tab:radial_grad}
   \begin{tabular}{c c c c c c}
        \hline
        &        &      & \multicolumn{2}{c}{NGC\,300} & \\
        & Object & Lines& $m$ & $n$ & Reference \\
        \hline
        12 + log(O/H) & {\hii} & RLs  & $-$0.14 $\pm$ 0.18 & 8.71 $\pm$ 0.10 & This work \\
                      & {\hii} & CELs & $-$0.30 $\pm$ 0.08 & 8.46 $\pm$ 0.05 & This work \\
                      & {\hii} & CELs & $-$0.41 $\pm$ 0.03 & 8.57 $\pm$ 0.02 & \citet{2009ApJ...700..309B} \\
                      & {\hii} & CELs & $-$0.36 $\pm$ 0.05 & 8.48 $\pm$ 0.03 & Stasi{\'n}ska et al.(2013) \\
                      & PNe    & CELs & $-$0.13 $\pm$ 0.08 & 8.35 $\pm$ 0.04 & Stasi{\'n}ska et al.(2013) \\
                      & B stars&      & $-$0.32 $\pm$ 0.26 & 8.58 $\pm$ 0.13 & \citet{2005ApJ...622..862U} \\
        12 + log(C/H) & {\hii} & RLs  & $-$0.43 $\pm$ 0.29 & 8.45 $\pm$ 0.16 & This work \\
        log(C/O)      & {\hii} & RLs  & $-$0.29 $\pm$ 0.26 & $-$0.26 $\pm$ 0.19 & This work \\
       12 + log(N/H)  & {\hii} & CELs & $-$0.83 $\pm$ 0.09 & 7.61 $\pm$ 0.05 & This work. ICF by Izotov et al. (2006) \\
                      &        & CELs & $-$0.82 $\pm$ 0.09 & 7.55 $\pm$ 0.06 & This work. Standard ICF \\
        log(N/O)      & {\hii} & CELs & $-$0.59 $\pm$ 0.16 & $-$0.82 $\pm$ 0.09 & This work. ICF by Izotov et al. (2006) \\
        $[ Z ]$       & A stars&      & $-$0.44 $\pm$ 0.12 & $-$0.06 $\pm$ 0.09 & \citet{2008ApJ...681..269K} \\
        \hline
         &        &      & \multicolumn{2}{c}{M\,33} & \\
                              & Object  & Lines&     $m$            &        $n$      & Reference \\
        \hline
        12 + log(O/H) & {\hii}  & RLs  & $-$0.33 $\pm$ 0.13 & 8.76 $\pm$ 0.07 & This work \\
                      & {\hii}  & CELs & $-$0.36 $\pm$ 0.07 & 8.52 $\pm$ 0.03 & This work \\
                      & {\hii}  & CELs & $-$0.36 $\pm$ 0.07 & 8.53 $\pm$ 0.05 & \citet{2007AA...470..865M} \\
                      & {\hii}  & CELs & $-$0.29 $\pm$ 0.07 & 8.48 $\pm$ 0.04 & \citet{2011ApJ...730..129B} \\
                      & {\hii}  & CELs & $-$0.18 $\pm$ 0.08 & 8.36 $\pm$ 0.04 & \citet{2008ApJ...675.1213R} \\
                      & PNe     & CELs & $-$0.21 $\pm$ 0.08 & 8.44 $\pm$ 0.06 & \citet{2009ApJ...696..729M} \\
                      & B stars &      & $-$0.41 $\pm$ 0.14 & 8.81 $\pm$ 0.11 & \citet{2005ApJ...635..311U} \\
        12 + log(C/H) & {\hii}  & RLs  & $-$0.61 $\pm$ 0.11 & 8.64 $\pm$ 0.18 & This work \\
        log(C/O)      & {\hii}  & RLs  & $-$0.28 $\pm$ 0.26 & $-$0.07 $\pm$ 0.16 & This work \\
        12 + log(N/H) & {\hii}  & CELs & $-$0.82 $\pm$ 0.13 & 7.69 $\pm$ 0.08 & This work. ICF by Izotov et al. (2006) \\
                      &         & CELs & $-$0.68 $\pm$ 0.13 & 7.59 $\pm$ 0.09 & This work. Standard ICF \\
        log(N/O)      & {\hii}  & CELs & $-$0.50 $\pm$ 0.08 & $-$0.80 $\pm$ 0.06 & This work. ICF by Izotov et al. (2006) \\       \hline
        
   \end{tabular}
  \end{minipage}
\end{table*}

\begin{figure*}
 \centering
  \includegraphics[width=0.49\textwidth]{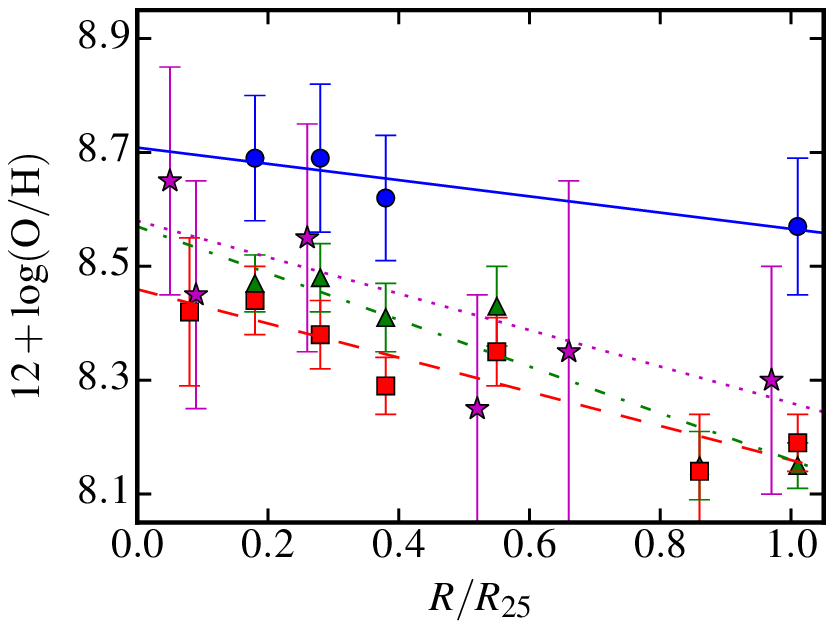}
  \includegraphics[width=0.49\textwidth]{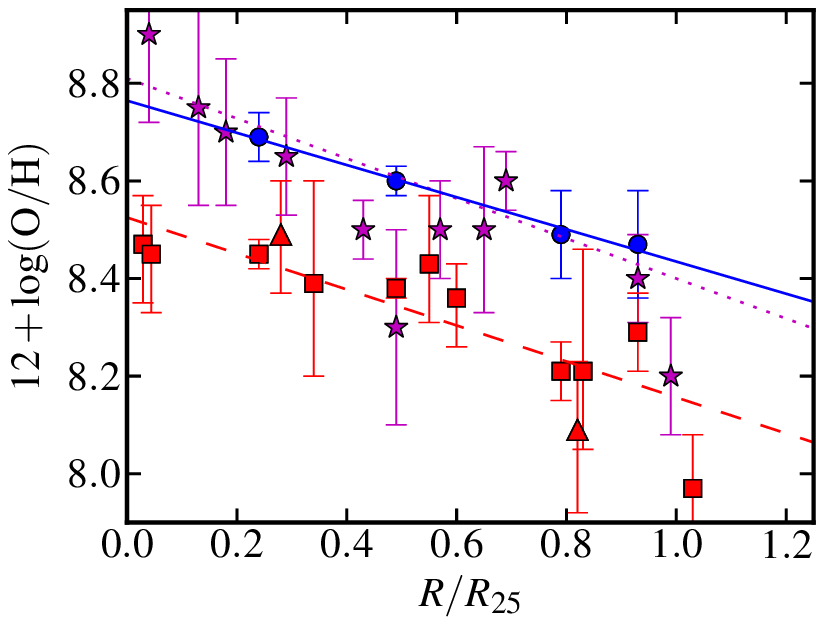}
  \caption{Comparison between the O/H ratios and radial gradients determined from our data for \hii\ regions sample and for other objects and authors. Left panel: NGC\,300, comparison of our data  (CELs: red squares and dashed line; RLs: blue circles and continuous line) with the results obtained by \citet{2009ApJ...700..309B} -- only for the \hii\ regions in common -- (green triangles and dashed-dotted line) and by \citet{2005ApJ...622..862U} 
  for B  supergiants (magenta stars and dotted line). Right panel: M\,33, comparison of our data (CELs: red squares and triangles and dashed line, RLs: blue circles and continuous line) with the result by \citet{2005ApJ...635..311U} for B supergiants (magenta stars and dotted line). The galactocentric distance of \citet{2005ApJ...635..311U} were recalculated adopting the isophotal radius of 28\arcmin, $\mathrm{i} = 53^{\mathrm{o}}$ and $\mathrm{P.A.} = 22^{\mathrm{o}}$.} 
  \label{fig:o_cels}
\end{figure*}

\section{Discussion}\label{sec:discusion}

\subsection{The radial O abundance gradient}\label{sec:o_gradient}

Abundance gradients are essential observational constraints for chemical evolution models of galaxies. \citet{1979MNRAS.189...95P}, \citet{1983MNRAS.204..743W}, \citet{1984MNRAS.211..507E}, \citet{1988A&AS...73..407D} and \citet{1997A&A...322...41C} determined abundances from the spectra of  {\hii} regions of NGC~300. However, in those papers, only two objects, one each in \citet{1979MNRAS.189...95P} and \citet{1983MNRAS.204..743W} had measurements of the auroral {\foiii} 4363 \AA\ line, which is necessary to determine $T_\mathrm{e}$. A substantial improvement was made by \citet{2009ApJ...700..309B}, who derived radial abundance gradients of He, N, O, Ne, S and Ar of NGC\,300 using direct determinations of $T_\mathrm{e}$ in 28 {\hii} regions. The gradients of those same elements were also determined by \citet{2013A&A...552A..12S} using deep spectra of 26 PNe and 9 compact {\hii} regions with direct determinations of $T_\mathrm{e}$. 
Additional studies about the gradients of NGC\,300 are based on detailed spectral analysis of supergiant stars. \citet{2005ApJ...622..862U} derived abundances of C, N, O, Mg and Si in six B supergiants at different galactocentric distances, allowing to compare with {\hii} region results. Finally, \citet{2008ApJ...681..269K} determined metallicities -- not elemental abundances -- for a sample of 24 A-type supergiants,  deriving the metallicity gradient across the disc of NGC\,300. 

Chemical abundances of \hii\ regions in M\,33 have been studied for four decades. \citet{1971ApJ...168..327S}, \citet{1975ApJ...199..591S}, \citet{1981MNRAS.195..939K}, \citet{1988MNRAS.235..633V}, and \citet{1997ApJ...489...63G} studied its chemical composition from the spectra of \hii\ regions.  \citet{2006ApJ...637..741C} were the first to derive O/H and Ne/H gradients with direct determination of the $T_\mathrm{e}$ in 13 \hii\ regions. \citet{2008MNRAS.387...45R} determined the radial gradients of Ne/H and S/H from \textit{Spitzer} spectra of \hii\ regions in M\,33. Later works by \citet{2011ApJ...730..129B} and \citet{2008ApJ...675.1213R} increased this information with new measurements of the {\foiii} 4363 \AA\ line. \citet{2009ApJ...696..729M} and \citet{2004A&A...426..779M} determined radial abundance gradients of M\,33 using PNe. \citet{1995ApJ...455L.135M} and \citet{1998ASPC..147...54V} derived chemical abundances from the spectra of two A-type supergiant stars. \citet{2005ApJ...635..311U} using B-type supergiants stars derived abundances for several chemical elements such as O, C, Ni, Mg and Si and computed radial abundance gradients.

We have performed least-squares linear fits to the fractional galactocentric distance of the objects, $R$/$R_{25}$ -- given in Table~\ref{tab:hii_data} -- and their O abundances included in Tables~\ref{tab:ion_abun}, \ref{tab:ion_abun_m33} and \ref{tab:abun_RLs}. These fits give the following radial O abundance gradients considering whether $\mathrm{O^{2+}}$/H$^+$ is determined either from CELs or RLs.
For NGC\,300:
 
\begin{equation}
12 + \log(\mathrm{O/H})_\mathrm{CELs} =8.46(\pm 0.05) - 0.30(\pm 0.08) R/R_{25};
\end{equation}

\begin{equation}
12 + \log(\mathrm{O/H})_\mathrm{RLs} =8.71(\pm 0.10) - 0.14(\pm 0.18) R/R_{25}.
\end{equation}
For M\,33, we have included the O/H ratios of the two \hii\ regions observed by \citet{2009ApJ...700..654E}:
\begin{equation}
12 + \log(\mathrm{O/H})_\mathrm{CELs} =8.52(\pm 0.03) - 0.36(\pm 0.07) R/R_{25};
\end{equation}

\begin{equation}
12 + \log(\mathrm{O/H})_\mathrm{RLs} =8.76(\pm 0.07) - 0.33(\pm 0.13) R/R_{25}.
\end{equation}

We compare graphically these O/H gradients in Figure~\ref{fig:DA} (NGC\,300: left panel; M\,33: right panel). The offset between both determinations of O/H for the same object and the different intercepts of the least-squares fittings illustrate the effect of the abundance discrepancy problem (see Section~\ref{t2}). The slopes of the fits 
obtained from CELs and RLs are somewhat different -- but marginally consistent within the errors -- in NGC\,300 and almost identical in the case of M\,33. O/H gradients based on RLs and CELs have also been derived for other few spiral galaxies (see Table~\ref{tab: summary}). In the case of the Milky Way, \citet{2005ApJ...618L..95E} find almost coincident slopes for the O/H gradients whether they use CELs and RLs. \cite{2009ApJ...700..654E} also report the same coincidence in data for the spiral galaxy M101, although the gradient derived by \citet{2003ApJ...591..801K} from CELs is somewhat steeper. 
Table~\ref{tab:radial_grad} shows least-squares fittings of O/H, C/H and C/O ratios with respect to $R$/$R_{25}$ -- radial gradients -- 
obtained for NGC~300 and M\,33 in this and previous works. In the table, we give the slope ($m$) and the intercept ($n$) of each fitting represented by the equation: 

\begin{equation}
12 + \log(\mathrm{X/H}), \log(\mathrm{X/Y}) = n + m \times R/R_{25}.
\end{equation}

\begin{figure*}
  \includegraphics[width=0.49\textwidth]{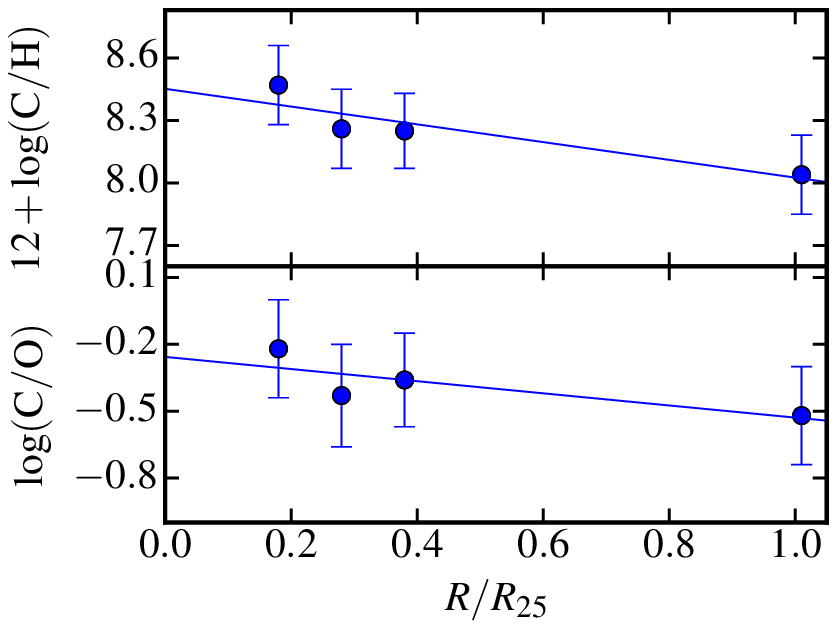}
  \includegraphics[width=0.49\textwidth]{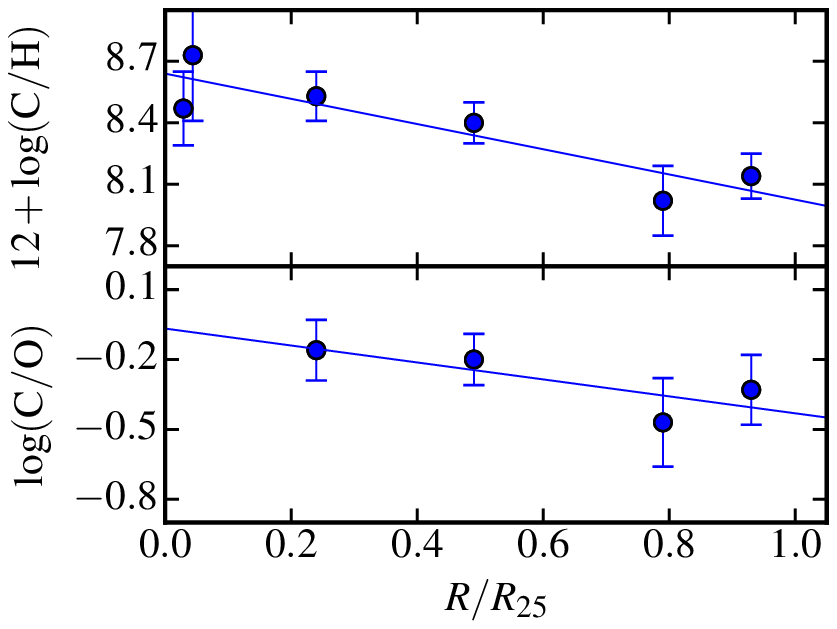}
  \caption{C/H (upper panel) and C/O (lower panel) gradients as a function of the fractional galactocentric distance, $R$/$R_{25}$, for NGC\,300 (left panel) and M\,33 (right panel).} 
  \label{fig:c_co}
\end{figure*}

In Table~\ref{tab:radial_grad} we see that the slope of the radial O/H gradient we determine from CELs in NGC\,300 is somewhat shallower than those obtained by \citet{2009ApJ...700..309B} and \citet{2013A&A...552A..12S}  from {\hii} regions -- but consistent within the errors with the second reference -- and in agreement with that determined by \citet{2005ApJ...622..862U} from B supergiants. The intercept of our gradient determined from CELs is similar to that given by \citet{2013A&A...552A..12S} but about 0.12 dex lower than the intercept obtained by \citet{2009ApJ...700..309B} and \citet{2005ApJ...622..862U}. Our O/H gradient determined from RLs is flatter than all the estimates based on CELs of {\hii} regions. The intercept of the O/H gradient derived from RLs is between 0.2 and 0.3 dex higher that the ones based on CELs due to the abundance discrepancy problem. The slope of the O/H gradient determined by \citet{2013A&A...552A..12S} for PNe is shallower that the ones obtained from CELs of {\hii} regions. They argue that this may be due to the steepening of the metallicity gradient in NGC~300 during the last Gyr, although there are suspicions that O in PNe may be altered by nucleosynthesis in their progenitors \citep[e.g.][]{2014PASA...31...30K, 2015MNRAS.449.1797D}.

 Table~\ref{tab:radial_grad} also includes the radial abundance gradients for M\,33. The slopes of our radial O/H gradients derived from CELs and RLs are almost identical and in agreement with that determined by \citet{2007AA...470..865M} for \hii\ regions, but the slopes obtained by \citet{2008ApJ...675.1213R},  \citet{2009ApJ...696..729M} and \citet{2011ApJ...730..129B}  are flatter than our determination but -- most of them -- consistent within the errors. The slope of the O/H gradient from B supergiants obtained by \citet{2005ApJ...635..311U} is the most steeper one in our comparison sample, but still consistent within the errors. The intercept of our gradient determined from CELs is similar to that calculated by \citet{2007AA...470..865M} and about 0.07 dex higher that those determined by \citet{2011ApJ...730..129B} and \citet{2009ApJ...696..729M}. As in NGC\,300, the intercept of the O/H gradients derived form CELs is around 0.25 dex lower that the ones based on RLs. The intercept calculated from B supergiants by \citet{2005ApJ...635..311U} is slightly higher than the one determined from RLs in \hii\ regions.

In Figure~\ref{fig:o_cels}, we represent the same information included in Figure~\ref{fig:DA} plus additional data from the literature. For NGC\,300 (left panel) we included: a) the O/H ratios determined by \citet{2009ApJ...700..309B} for the {\hii} regions in common with us (green triangles); b) the O abundance gradient calculated by \citet{2009ApJ...700..309B} for all their observational points (green dashed-dotted line); c) the O/H ratios determined by \citet{2005ApJ...622..862U} for B supergiants (purple dotted stars) and d) the O abundance gradient determined by \citet{2005ApJ...622..862U} (purple line). Figure~\ref{fig:o_cels} indicates that the O/H ratios we determine from CELs are in good agreement with those obtained by \citet{2009ApJ...700..309B} for the same objects and with the O abundances calculated for B supergiants in NGC\,300 galaxy. Moreover, although not included in Figure~\ref{fig:o_cels}, the O/H ratios 
and gradients determined for {\hii} regions and PNe by \citet{2013A&A...552A..12S} are also in general agreement with the nebular values determined from CELs and stellar ones included in Figure~\ref{fig:o_cels}. The additional information included in Figure~\ref{fig:o_cels} for M\,33 is: a) the O/H ratios determined by \citet{2005ApJ...635..311U} for B supergiants (purple dotted stars) and b) the O abundance gradient determined by \citet{2005ApJ...635..311U} (purple line). The galactocentric distances of \citet{2005ApJ...635..311U} for B supergiants were recalculated adopting the isophotal radius of 28\arcmin, $\mathrm{i} = 53^{\mathrm{o}}$ and $\mathrm{P.A.} = 22^{\mathrm{o}}$, the values we have assumed in this paper. We discuss these results in \ref{sec:nebular_vs_stellar}.

\subsection{Comparison between nebular and stellar O abundances}\label{sec:nebular_vs_stellar}

One of the aims of this paper is to derive O abundances from RLs of {\hii} regions in NGC\,300 and M\,33 and to compare with determinations based on CELs for the same objects and abundances determined for B stars. 
As it is stated in Section~\ref{sec:o_gradient}, the values of O/H ratios based on RLs that we have derived are systematically between 0.2 and 0.4 dex higher than those based on CELs. 
In the case of NGC~300, the O abundances calculated by \citet{2005ApJ...622..862U} for B supergiants are consistent with our nebular values based on CELs as it can be seen in Figure~\ref{fig:o_cels}. This situation contrasts with the results obtained for M33. In this case 
our nebular O/H ratios based on RLs are more consistent with the quantitative spectroscopical analysis of B supergiants performed by \citet{2005ApJ...635..311U}. As it is shown in Figure~\ref{fig:o_cels}, both sets of data give O/H ratios about 0.3 dex larger than our determinations based on CELs and the results obtained by \citet{2007AA...470..865M}.  

It must be remarked that some fraction of O in nebulae may be embedded in dust grains and this should be considered when comparing with stellar abundances. \cite{2010ApJ...724..791P} have estimated that the depletion of O increases with increasing O/H. They propose that O depletion ranges from about 0.08 dex, for the metal poorest {\hii} regions, to about 0.12 dex -- the value estimated by \cite{2009MNRAS.395..855M} for the Orion Nebula -- for metal-rich ones. As we can see, the expected amount of O depleted in {\hii} regions 
should be about 0.1 dex, of the order of the typical uncertainties of our O abundance determinations. Therefore, the effect of dust depletion 
cannot account for the aforementioned 0.3 dex difference between the nebular O/H ratios determined from CELs and the stellar ones in M\,33.

It is important to note that O/H ratios determined from nebular CELs are always lower than stellar determinations in other galaxies where both kinds of data can be compared. In the case of M81, there is an offset between CELs-based nebular O/H ratios and stellar ones 
\citep{2012MNRAS.422..401P, 2012ApJ...747...15K}. Similar differences have been reported in M31; the nebular direct O abundance determinations based on CELs obtained by \citet{2012MNRAS.427.1463Z} are about 0.3 dex lower than the O/H ratios derived from B supergiants by \citet{2002A&A...395..519T} and \citet{2001MNRAS.325..257S} and from AF stars studied by \citet{2000ApJ...541..610V}. There is a single determination of O/H ratios from RLs in an {\hii} region of M31 \citep[object K932,][]{2009ApJ...700..654E}. These authors found that the O/H ratio derived from RLs for that object is about 0.2 dex higher than that obtained from CELs, being the RL determination more consistent with the stellar abundances extrapolated at the same galactocentric distance. 
 
For the Milky Way there are several works in the literature that compare nebular and B-type star O abundances. \citet{2011A&A...526A..48S} compare the results of gas+dust abundances of the Orion Nebula with stellar abundances of a sample of 13 B-type stars from the Orion star-forming region (Ori OB1). These authors find that the O/H ratios based on RLs agree much better with the stellar abundances than the ones  derived from CELs. Moreover, \citet{2012A&A...539A.143N} have established 
a present-day cosmic abundance standard from a sample of 29 early B-type stars, finding that the O/H ratio of the solar vicinity is more consistent with the gas+dust abundance of the Orion Nebula determined by \citet{2004MNRAS.355..229E} with RLs. In fact, assuming the CELs values of the Orion Nebula as representative of the current O/H ratio of the local ISM would imply an unrealistic large fraction of O depleted onto dust. 
Finally, the results obtained by \citet{2014A&A...571A..93G} for the Cocoon nebula point in the same direction. The comparison of the O and N abundances determined for the ionizing B0.5 V star are higher than the nebular 
gas+dust abundances determined from CELs; \cii\ and \oii\ RLs cannot be detected in this object due to its low ionization degree. In the Cocoon nebula, stellar and nebular O abundances would only agree assuming a moderate abundance discrepancy 
factor of the order of that estimated for the Orion Nebula. 

As we can see from the discussion above and in the context of 
the paucity of observational data available, NGC~300 stands as the only galaxy where the O abundances determined from CELs of {\hii} region spectra are more consistent with those determined from early-type stars. In any case, both the stellar and RLs-based determinations of the O abundance have rather large uncertainties in the objects of this galaxy and the offset between both kinds of determinations can even be regarded as marginal. 

\subsection{The radial C abundance gradient}\label{sec:c_gradient}

\begin{table*} 
\caption{Comparison of O, C, C/O, N and N/O gradients for several spiral galaxies.} 
\center
\begin{tabular}{ccccccccccc} 
\hline\hline 
           &              &          &                 &                    \multicolumn{7}{c}{slope(dex ($R$/$R_{25})^{-1}$)}                                       \\
           & Morphological& $R_{25}$ &                 & \multicolumn{2}{c}{$\mathrm{O/H}$$^{\rm a}$}         & $\mathrm{C/H}$& \multicolumn{2}{c}{$\mathrm{C/O}$$^{\rm a}$}    & $\mathrm{N/H}$     & $\mathrm{N/O}$   \\
Galaxy     &  type        &  (kpc)   & $M_{\mathrm V}$ &        (RLs)    &     (CELs)               & (RLs)         &  (RLs/RLs)     &  (RLs/CELs)       & (CELs)             & (CELs)            \\
\hline                                                                                                                       &                                                                    \\
NGC~300    & Sc           & 5.33     & $-$18.99        & \textbf{$-$0.14}         & $-$0.30         & $-$0.43       &  \textbf{$-$0.29}        & $-$0.13 & $-$0.83            & $-$0.59           \\
M33        & SAcd         & 6.85     & $-$19.41        & \textbf{$-$0.33}& $-$0.36                  & $-$0.61       &\textbf{$-$0.28}&  $-$0.25          & $-$0.82            & $-$0.50           \\
NGC~2403   & SAcd         & 7.95     & $-$19.51        & --              &\textbf{$-$0.32$^{\rm b}$}& $-$0.77       &  --            &  \textbf{$-$0.45} & $-$0.60$^{\rm b}$  & $-$0.37$^{\rm b}$ \\
NGC~628    & SAc          & 10.95    & $-$20.77        &  --             & $-$0.49$^{\rm c}$        &  --           &  --            &    --             & $-$1.33$^{\rm c}$  & $-$0.85$^{\rm c}$ \\
Milky Way  & SBbc         & 11.25    & $-$20.90        & \textbf{$-$0.41}& $-$0.48$^{\rm d}$        & $-$0.90       &\textbf{$-$0.49}&  $-$0.42          & $-$1.23$^{\rm e}$  & $-$0.89$^{\rm e}$ \\
M51        & SAbc         & 12.89    & $-$20.94        &  --             & $-$0.30$^{\rm f}$        &  --           &  --            &   --              & $-$0.92$^{\rm f}$  & $-$0.62$^{\rm f}$ \\
M101       & SABc         & 28.95    & $-$21.36        & \textbf{$-$0.50}& $-$0.90$^{\rm g}$        & $-$1.30       &\textbf{$-$0.80}&  $-$0.40          & $-$1.29$^{\rm g}$  & $-$0.35$^{\rm g}$ \\
M31        & SAb          & 21.75    &$-$21.78         & --              &\textbf{$-$0.48$^{\rm h}$}& $-$1.72       &  --            & \textbf{$-$1.24}  & $-$0.72$^{\rm h}$  & $-$0.29$^{\rm h}$ \\
\hline

           \multicolumn{10}{l}{$^{\rm a}$ The boldface values of O/H and C/O gradient slopes are those used in Figure \ref{fig:slope}.} \\
           \multicolumn{10}{l}{$^{\rm b}$ \citet{2013ApJ...775..128B}.}\\
           \multicolumn{10}{l}{$^{\rm c}$ \citet{2015ApJ...806...16B}.}\\
           \multicolumn{10}{l}{$^{\rm d}$ \citet{2015MNRAS.452.1553E}.}  \\
           \multicolumn{10}{l}{$^{\rm e}$ \citet{2005ApJ...623..213C}.}\\
           \multicolumn{10}{l}{$^{\rm f}$ \citet{2015ApJ...808...42C}.}\\           
           \multicolumn{10}{l}{$^{\rm g}$ \citet{2003ApJ...591..801K}. Gradients correspond to the inner disc. \citet{2016MNRAS.tmp...29P} find a break of the N/H }\\ 
           \multicolumn{10}{l}{gradient at $R/R_{G} = 0.7$, this has not been considered in our analysis.}\\
           \multicolumn{10}{l}{$^{\rm h}$ \citet{2012MNRAS.427.1463Z}.}\\

\end{tabular} 
\label{tab: summary} 
\end{table*}

\begin{figure}
\includegraphics{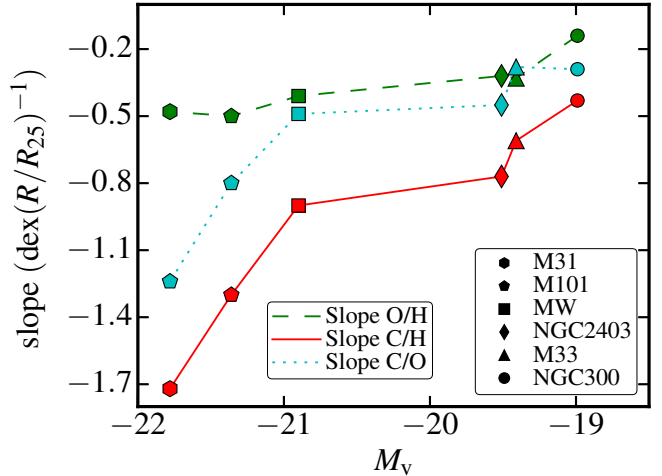}
\caption{Slope of O/H (green), C/H (red) and C/O (cyan) radial gradients $vs.$ absolute magnitude, $M_\mathrm{V}$ of several spiral galaxies: NGC~300 (squares), M33 (pentagons), NGC2403 (hexagons), Milky Way (circles), M101 (triangles) and M31 (diamonds). The values of the slope of O/H and C/O represented in this figure are those highlighted in boldface in Table \ref{tab: summary}.}
\label{fig:slope}
\end{figure}

Compared to what happens with O, the shape of the C abundance gradient has never been explored before in NGC\,300, and only by 
\citet{2009ApJ...700..654E} in the case of M\,33 -- but based on measurements for only two \hii\ regions,  NGC\,595 and NGC\,604. \citet{2005ApJ...622..862U} and \citet{2005ApJ...635..311U} estimated C/H ratios in B supergiants in NGC\,300 and M\,33. These authors did not explore the C gradient because their abundances, based on the stellar \cii\ 4267 \AA\ line, yield unrealistic low abundances when compared with other indicators. In addition, they found evidences of CNO-cycle contamination in the C and N abundances of their targets in NGC\,300. 

In Figure~\ref{fig:c_co}, we plot the C/H and C/O ratios as a function of the fractional galactocentric distance ($R/R_{25}$). We include the data for the aforementioned two \hii\ regions studied by \citet{2009ApJ...700..654E}. 
In the figure, we can see that the slope of the C/H gradient is steeper than that of O/H, producing a negative gradient of 
the C/O ratio in both galaxies. This is also the behavior found in other nearby spiral galaxies. In Table~\ref{tab: summary}, we summarize the slope of the O, C and C/O gradients we obtain for NGC\,300, M\,33 and previous results obtained by our group for the Milky Way \citep{2013MNRAS.433..382E} and other nearby spirals: NGC~2403, M101 \citep{2009ApJ...700..654E} and M31 \citep{2014AN....335...73E}.  In addition, we also give information about the morphological type, the 25th magnitude $B$-band isophotal radius, $R_{25}$, and absolute V-band magnitude, $M_\mathrm{V}$, of the galaxies. 
In Figure~\ref{fig:slope}, we represent the slope of O/H, C/H and C/O radial gradients (discontinuous green, continuous red and dotted cyan lines, respectively) with respect to the absolute magnitude, $M_{\rm V}$, of the galaxies 
included in Table~\ref{tab: summary}. For deriving the C/O gradients represented in Figure~\ref{fig:slope}, we have used the O/H slopes obtained from RLs in those galaxies where 
these kinds of lines have been detected, for the rest -- NGC~2403 and M31 -- we used the O/H slopes obtained from CELs. Figure~\ref{fig:slope} indicates that since the slope of the O/H gradient shows a slight correlation with $M_{\rm V}$ -- that is in contradiction with recent results of studies based on the application of strong-line methods for deriving metallicities in large samples of 
spiral galaxies \citep[e.g.][]{2014A&A...563A..49S, 2015MNRAS.448.2030H} -- the slope of the C/H gradient shows a strong correlation with $M_{\rm V}$,  
i.e. the Milky Way, M31 and M101 galaxies show systematically steeper C/H gradients than NGC~300, M33 and NGC~2403, which are less luminous and less massive spirals. Moreover, the rate at which the slope becomes steeper 
increases as the galaxy is more luminous or massive. As expected, the C/O ratio also shows a clear correlation. Summarizing, the results we show in Table~\ref{tab: summary} and Figure~\ref{fig:slope} indicate that  the slope of C/H is always steeper than that of O/H in the galaxies where gradients of both 
elements have been estimated. This result has to be further investigated because the number of {\hii} regions with these kinds of data available is still very small in most of the galaxies, specially in the cases of M\,31 and NGC~2403, where only two or three observational points are available. Therefore, it is necessary to detect \cii\ and \oii\ RLs in more {\hii} regions in the discs of nearby spiral galaxies in order to better characterize this correlation.

The origin of negative slopes of C/O gradient has been explored by \citet{2005ApJ...623..213C}. Based on chemical evolution models of the Milky Way, those authors found that they can be reproduced when considering C yields that increase with metallicity owing to stellar winds in massive stars and decrease with metallicity owing to stellar winds in low- and intermediate- mass stars. As a result of that work, at high metallicities, 12+log(O/H) $\geq$ 8.5, the main contributors to C should be massive stars. However, at intermediate metallicities,  8.1 $\leq$ 12+log(O/H) $\leq$ 8.5, the main contributors to C are low metallicity low-mass stars. The behaviour of the C/O radial abundances gradients are consistent with an ``inside-out'' formation scenario of inner parts of spiral galaxy discs \citep[see e.g.][model for M\,33]{ 2013MNRAS.429.2351R}. In the inner parts of galaxies, the low- and intermediate-mass stars had enough time to inject C into the ISM. The newborn massive stars formed in more metal-rich environments can further enrich the ISM via their stellar winds. Therefore, this behavior would certainly produce different C/H gradients in galaxies of different luminosities and average metallicities. In upcoming papers we will further explore the theoretical reasons of the variations of the slope of the C/O gradients depending on galactic properties. 

\begin{figure}
   \centering
   \includegraphics[scale=0.44]{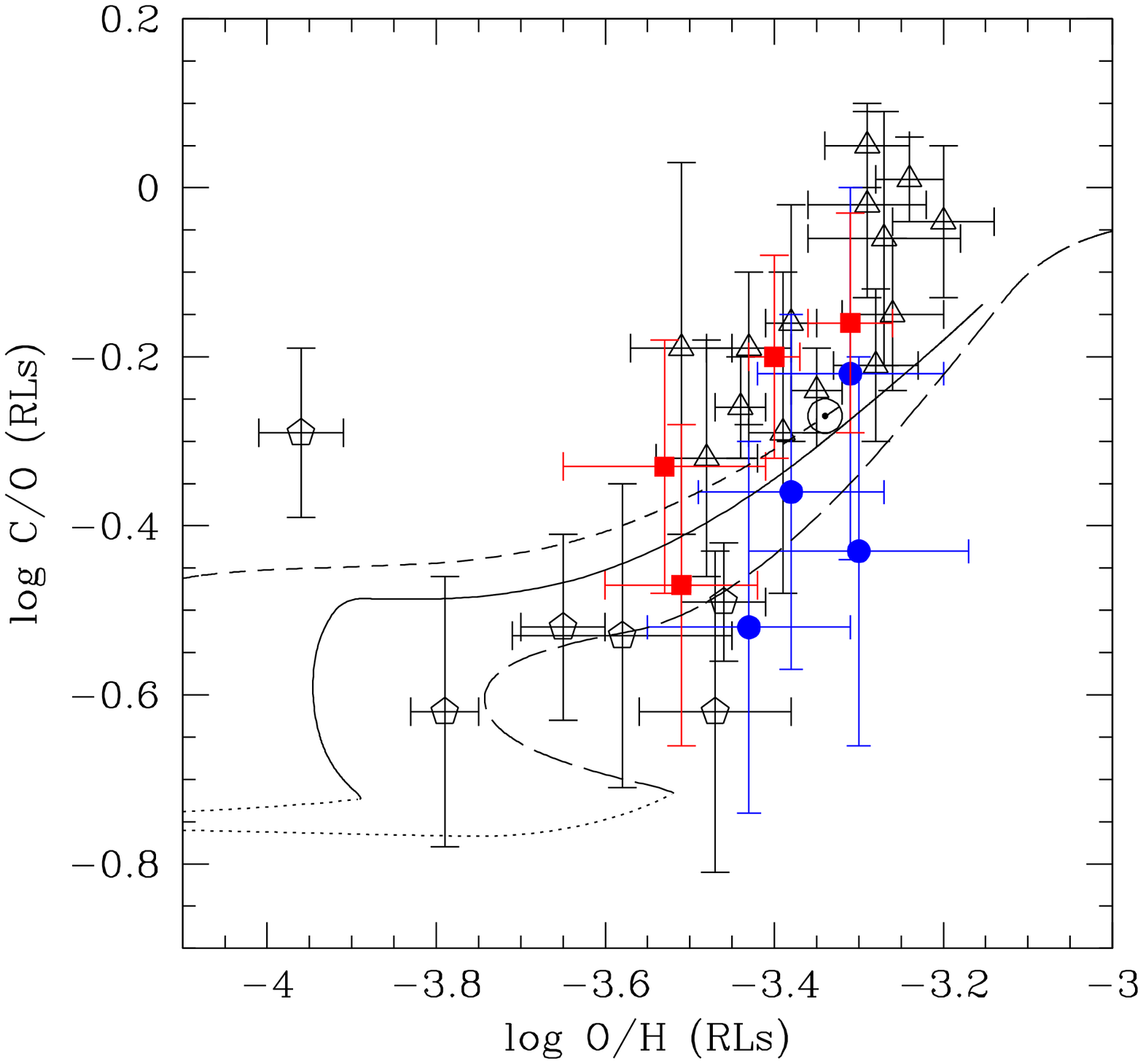} 
   \caption{C/O {\it vs.}~O/H ratios in  \hii\ regions determined from recombination lines. Red full rectangles correspond to 
   objects in M\,33 and blue full circles represent \hii\ regions in NGC~300. We also include data for \hii\ regions in the Milky Way \citep{2007ApJ...670..457G, 2013MNRAS.433..382E} and in the inner discs of external spiral galaxies \citep{2009ApJ...700..654E} -- open triangles -- and star-forming dwarf galaxies 
   \citep[see][an references therein]{2014MNRAS.443..624E} -- open pentagons. The solar symbol 
  represents the abundances of the Sun \citep{2009ARA&A..47..481A}. The lines show the predictions of chemical evolution models presented by \citet{2013MNRAS.433..382E} 
  for the Milky Way. The solid line represents the time evolution of abundances since the formation of the Galactic disc ($t_G$ $>$ 1 Gy) for $R_G$ = 8 kpc, the long- and short-dashed lines represent the same but for $R_G$ = 4 and 14 kpc, respectively. Dotted lines represent the evolution for the same radii but for $t_G$ $<$ 1 Gy, during the formation of the Galactic halo.  We recommend to add $\sim$0.1 dex to the log(O/H) values of {\hii} regions to correct for dust depletion. No correction is needed for their C/O ratios \citep[see][for details]{2014MNRAS.443..624E}. }
   \label{fig:covso}
  \end{figure}

Figure~\ref{fig:covso} plots the C/O {\it vs.}~O/H ratios of the  {\hii} regions of NGC~300 and M\,33 included in Table~\ref{tab:abun_RLs} and the two objects of M\,33 studied by \citet{2009ApJ...700..654E} --those for which we have abundance determinations from RLs in both galaxies -- as well as similar data for {\hii} regions in the Milky Way and other spiral galaxies and star-forming dwarves. 
We can see that most of the nebulae of NGC~300 and M\,33 show C/O ratios similar to {\hii} regions belonging to the inner discs of other spiral galaxies. However, two of the most external objects: R2 (in NGC~300, $R$/$R_{25}$ = 1.01) and NGC~588 (in M\,33, $R$/$R_{25}$ = 0.79), show lower values of the C/O ratio, similar to those  typical of {\hii} regions in star-forming dwarf galaxies. This result contrasts that obtained for NGC~2579, an {\hii} region of the Milky Way located at $R$/$R_{25}$ = 1.1, for which \citet{2013MNRAS.433..382E} obtain a log(C/O) = $-$0.26, exactly the solar value. The so different C/O ratios in R2 and NGC~588 with respect to NGC~2579 suggests different chemical histories in the external zones of the Milky Way and the small spiral galaxies NGC~300 and M\,33. \citet{2013MNRAS.433..382E} found that O/H, C/H and C/O gradients seem to flatten in the outer disc of the Milky Way, but no evidences of such flattening have been found in the external disc of NGC~300 \citep[e.g.][]{2009ApJ...700..309B, 2013A&A...552A..12S} nor in M\,33 \citep[e.g.][]{2007AA...470..865M}. It would be interesting to explore the behavior of the C/O gradients in the external zones of other similar low-mass spiral galaxies to ascertain whether this is a general rule or not. 

\subsection{The radial N abundance gradient}\label{sec:n_gradient}

\begin{figure}
  \includegraphics[width=0.47\textwidth]{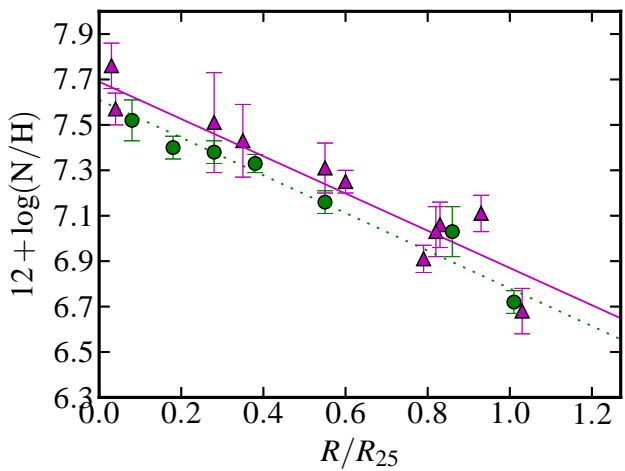}
  \caption{N/H ratios and gradient as a function of the fractional galactocentric distance, $R$/$R_{25}$, for {\hii} regions in NGC\,300 (green circles and dotted line) and M\,33 (magenta triangles and continuous line). The N/H ratios have been derived assuming the ICF scheme by \citet{2006A&A...448..955I}.} 
  \label{fig:n_gradient}
\end{figure}

Although an ICF is needed to derive the total N abundance of {\hii} regions -- and therefore may be affected by large systematic uncertainties -- we find that N/H ratios delineate a fairly clear gradient in NGC~300 and M\,33. As it is indicated in Section~\ref{sec:total}, we have used the N$^+$ abundance determined from CELs and the ICF by \citet{2006A&A...448..955I} to derive the N abundance of each object. In 
Figure~\ref{fig:n_gradient} we show the N/H ratios as a function of the fractional galactocentric distance, $R/R_{25}$, for each galaxy. In Table~\ref{tab:radial_grad} we include the values of the slope and intercept of the least squares fits of the N/H and N/O ratios. It is remarkable that all parameters of the N/H and N/O gradients are very similar for NGC~300 and M\,33, specially the slope, which is almost identical. In order to be more confident of this result, we also have derived the N/H gradients making use of the common  ICF scheme based on the similarity between the ionization potential of N$^+$ and O$^+$ proposed by \citet{1969BOTT....5....3P}, finding rather consistent values in both cases and for both galaxies (see values in Table~\ref{tab:radial_grad}). The average value of the difference between N/H ratios with the two ICF schemes used is 0.05 dex in NGC\,300 and 0.03 dex in M\,33. The values obtained using the ICF by \citet{1969BOTT....5....3P} are always lower. In addition, the N/H gradients as a function of $R/R_{25}$ obtained from the data of \citet{2009ApJ...700..309B} for NGC~300 and \citet{2010A&A...512A..63M} for M\,33 are  $-$0.85 and $-$0.55 dex ($R$/$R_{25})^{-1}$, respectively, in excellent agreement with our determination in the first case and somewhat less steeper in the second. As it is well known, the bulk of N and O are produced by stars of different initial masses. The N abundance increases at a faster rate than the O abundance because of secondary nitrogen becomes dominant since $12+\log(\mathrm{O/H}) > 8.3$ \citep{2000ApJ...541..660H} so their enrichment timescales are different and this is reflected in the negative gradients of the N/O ratio. 

In Table~\ref{tab: summary} we present the N/H and N/O gradients of other galaxies for comparison, including those galaxies for which we have C/H ratio determinations -- NGC~2403, Milky Way, M101 and M31 -- but also for two additional nearby spirals, NGC~628 and M51, with high- quality direct abundance determinations 
by \citet{2015ApJ...806...16B} and \citet{2015ApJ...808...42C}, respectively. Inspecting the values of the table, there seems to be some correlation between N/H slopes and $M_{\mathrm V}$ resembling the aforementioned C/H vs. $M_{\mathrm V}$ correlation but with the odd behavior of M31. There is no clear correlation in the case of the N/O gradients. 

\section{Conclusions}\label{sec:conclusions}

We present deep spectrophotometry of seven {\hii} regions of NGC~300 and eleven of M\,33. These are two nearby low-mass spiral galaxies that host bright star-forming regions. The data of NGC~300 objects have been taken with UVES echelle spectrograph attached to the UT2 unit of the VLT at Cerro Paranal Observatory. The spectra of M\,33 {\hii} regions have been obtained with the low-intermediate resolution spectrograph OSIRIS -- using its long-slit mode -- at the 10.4m GTC telescope at Roque de los Muchachos Observatory. We have derived precise values of the physical conditions for each object making use of several emission line-intensity ratios as well as abundances for several ionic species from the intensity of CELs. It is remarkable that we have obtained direct determinations of the electron temperature in all the observed objects. We detect pure RLs of {\cii} and {\oii} in several of the {\hii} regions, permitting to derive their C/H and C/O ratios.

We have derived the radial abundance gradient of O for each galaxy making use of CELs and RLs, as well as the C and N gradients using RLs and CELs, respectively. In the case of M\,33, we find that the nebular O/H ratios and the gradient determined from RLs are more consistent with the values obtained for B stars that those determined from CELs. The opposite situation is found for NGC~300, the O/H ratios obtained from CELs are more consistent with stellar abundances. One important result of this work is the first determination of the C/H gradient of NGC~300 and an improving of its determination in the case of M\,33. In both galaxies, the C/H gradients are steeper that those of O/H, leading to negative C/O gradients. This result is consistent with the predictions of an ``inside-out'' formation scenario. Comparing with similar results for other spiral galaxies, we find a strong correlation between the slope of the C/H gradient and $M_\mathrm{V}$, in the sense that steeper C/H -- and C/O -- gradients are shown by most luminous galaxies. We find that some {\hii} regions located close to the $R_{25}$ of NGC~300 and M\,33 show C/O ratios more similar to those typical of dwarf galaxies than those of  {\hii} regions in the discs of more massive spirals. This may be related to the absence of flattening of the gradients in the external parts of NGC~300 and M\,33. Finally, we find very similar N/H gradients in both galaxies, we also find 
some correlation between the slope of the N/H gradient and $M_\mathrm{V}$ comparing with similar data for a sample of spiral galaxies. 

\section*{Acknowledgments}

This work is based on observations collected at the European Southern Observatory, Chile and the Gran Telescopio de Canarias (GTC), instaled in the Spanish Observatory del Roque de los Muchachos of the Instituto de Astrof\'isica de Canarias, in the island of La Palma. LTSC is supported by the FPI Program by the Ministerio de Econom\'ia y  Competividad  (MINECO) under grant AYA2011-22614. LTSC would like to thank Luis Peralta de Arriba for comments that helped
to improve the presentation of the results. JG-R acknowledges support from Severo Ochoa excellence program (SEV-2011-0187) postdoctoral fellowship.

\label{lastpage}

\end{document}